\documentclass[a4paper,12pt]{article}
\usepackage{epsfig}
\usepackage{graphicx}
\usepackage{dcolumn}
\usepackage{bm}
\usepackage{longtable}
\usepackage{amssymb}
\usepackage{amsfonts}
\usepackage[margin=1.0in]{geometry}
\usepackage{hyperref}
\usepackage{amsmath}
\allowdisplaybreaks[1]

\def\a{\alpha}
\def\b{\beta}
\def\g{\gamma}

\def\d{\delta}
\def\e{\epsilon}

\def\s{\sigma}

\def\om{\omega}

\def\p{\partial}
\def\half{{1\over 2}}

\def\m{\mu}
\def\n{\nu}

\makeatletter							  
\@addtoreset{equation}{section}					  
\makeatother							  

\begin{document}

\begin{titlepage}
\begin{flushright}
DFTT 14/2011\\
DISTA-2011
\par\end{flushright}
\vskip 1.5cm
\begin{center}
\textbf{\huge \bf  Supersymmetric Fluid Dynamics}
\textbf{\vspace{2cm}}\\
{\Large P.~A.~Grassi$ ^{~a,c,}$\footnote{pgrassi@mfn.unipmn.it}, 
A.~Mezzalira$ ^{~a,b,c,}$\footnote{mezzalir@to.infn.it}, and L. Sommovigo$ ^{~a,}$\footnote{lsommovi@mfn.unipmn.it}}

\begin{center}
{a) { \it DISTA, Universit\`{a} del Piemonte Orientale,
}}\\
{{ \it via T. Michel, 11, Alessandria, 15120, Italy }} \\ \vspace{.2cm}
 {b) { \it Dipartimento di Fisica Teorica, Universit\`a di Torino,}}\\
 {{\it via P. Giuria, 1, Torino, 10125, Italy }}
\\ \vspace{.2cm}
 {c) { \it INFN
- Sezione di Torino}}
\end{center}

\par\end{center}
\vfill{}

\begin{abstract}
{ \noindent
Recently Navier-Stokes (NS) equations have been derived from the duality between the black branes and a conformal fluid on the boundary of $AdS_5$. 
Nevertheless, the full correspondence has to be established between solutions of supergravity in $AdS_{5}$ and supersymmetric field theories on the boundary. That prompts the construction 
of NS equations for a supersymmetric fluid.
In the framework of rigid susy, there are several possibilities and we propose one candidate.
We deduce the equations of motion in two ways: both from the divergenless condition on the energy-momentum tensor and by a suitable parametrization of the auxiliary fields.
We give the complete component expansion and a very preliminary analysis of the physics of this supersymmetric fluid.
}
\end{abstract}
\vfill{}
\vspace{1.5cm}
\end{titlepage}

\vfill
\eject

\tableofcontents
\setcounter{footnote}{0}


\section{Introduction}

Recently, several attention has been given to the derivation of the relativistic/non-relativistic fluid-dynamics from a gravity dual theory. This is supported by well-known AdS/CFT correspondence which led, in this specific framework, to new important results \cite{Bhattacharyya:2008jc,Banerjee:2008th,Erdmenger:2008rm,Bhattacharyya:2008ji,Bhattacharyya:2008kq,Rangamani:2009xk,Fouxon:2008tb,Compere:2011dx,Bredberg:2011jq,Lysov:2011xx}. 
Nevertheless, the exact duality can only be established within the framework of supergravity 
on one side and supersymmetric (conformal) field theories on the dual side. In the present case, it means that the fluid on the 
boundary must be a supersymmetric one whose dynamics is governed by supersymmetric equations. 
Since  the bosonic dynamics is determined by a divergence-free current $j^\mu$ and a set of equations known as Navier-Stokes (NS) equations, we provide their supersymmetric extension and we discuss some implications.\footnote{A first step in that direction has been done in \cite{Gentile:2011jt}, where the corrections to NS eqs. due to fermionic 0-modes have been computed.} 

There are few accounts on the subject. We have to recall papers \cite{Nyawelo:2003bv,Nyawelo:2003bw} where a supersymmetric extension of the 
Navier-Stokes equations is provided.\footnote{Note that with the terminology 
superfluid is intended a quantum fluid in certain conditions where the quantum properties become relevant and it behaves in peculiar way 
\cite{KH}}
However, that extension is not the most suitable one for our purposes: there the
current $j^\mu$ is encoded into a real superfield, whilst it must be 
replaced by a linear superfield. Indeed, a real superfield forces a certain 
dynamics which, by setting $C$ and $\omega$ -- the superparteners of $j^\mu$ -- to 
zero, does not  reduce to a generic bosonic fluid. 
The equations obtained in papers \cite{Nyawelo:2003bv,Nyawelo:2003bw} 
substantially differ from ours. Another extension of the NS equations is provided in \cite{Jackiw:2004nm}. 
In that work, the authors start from a 3d model and supersymmetrize it by adding a pair of fermions. It is not clear to 
us how to perform the same construction for relativistic 4d model. 

Our construction is obtained as follows: first, we review the derivation of the NS equations in the 
bosonic framework. The action is given in terms of a function $f(j^2)$ whose form depends upon the equation 
of state of the fluid and whose argument is the square of $j^\mu$. The latter is also coupled to an auxiliary field 
$a_\mu$. 

We show that NS equations can be obtained in two ways: on the one hand, by computing the energy-momentum tensor and imposing the divergenceless condition. 
Doing so, the auxiliary 
field $a_\mu$ is replaced by its equation of motion and it disappears from NS equations. On the other hand, 
is observing that the curvature of the auxiliary field $a_\mu$ (viewed as a gauge field) contracted with the current $j^\mu$ leads to the same equations without computing the energy-momentum tensor. This is due to the fact that the action 
is invariant under certain isometries if and only if NS equations are satisfied. 
We analyze in detail the equivalence between the two approaches since it is relevant in the supersymmetric case. 

Still in purely bosonic framework, we discuss the importance of the so-called Clebsch paremeterization 
of the auxiliary field $a_\mu$ \cite{Jackiw:2004nm,Zyoshi:2010,clebsch}. According to Clebsch, $a_\mu$ is expressed in terms of some potentials 
whose equations of motion lead to NS eqs. The method is very straightforward and we will follow the same 
technique also in the susy case. As discussed in \cite{Nyawelo:2003bv,Nyawelo:2003bw}, there is 
another way to introduce the Clebsch potentials using a K\"ahler potential. 
We prove the equivalence with usual Clebsch parametrization and we discuss the relevance of that choice. 

To move on, we describe the set of superfields needed in the susy framework. 
The current $j^\mu$ is taken as the middle component of linear superfield $J$ which also contains a fundamental scalar field $C$, 
plus the fermionic field $\omega$. We recall that a linear superfield in 4d has the same 
number of d.o.f. as the chiral multiplet without auxiliary fields and the current, which appears as a $\theta^2$ component,
is conserved as a consequence of the linearity of $J$ \cite{West:1990tg,Wess:1992cp,Gates:1983nr,Weinberg:2000cr}. 

As already stated, the dynamics is characterized by a function $F(j^2)$ which determines also the equation of state. However, 
to reproduce the bosonic action, the argument of that function must be the current $j^\mu$ itself. Therefore a new linear superfield ${\cal J}_\mu$ 
(such that any value of the index $\mu$ labels a linear superfield), whose first component is the current $j_\mu$, has to be introduced. This is obtained by acting with 
superderivatives on $J$.  The auxiliary fields such as $a_\mu$ of the bosonic theory are now encoded into a real superfield $A$ and the 
complete supersymmetric Lagragian can be easily obtained.  

We provide the complete 
Lagrangian by expanding the superfields in components and integrating over the $\theta$'s. Due to 
this expansion, the number of possible terms increases and the Lagrangian is really cumbersome. For 
that, to grasp the meaning of it, we derive the superfield equations of motion and we compute their bosonic sector. The energy-momentum tensor for the Lagrangian restricted to the physical field $C$
is computed and some considerations are proposed.  

One important issue is the dependence of the K\"ahler potential. In the supersymmetric case 
the identification of the abelian real superfield $A$ with the real K\"ahler potential is 
rather clear (see also \cite{Binetruy:2000zx}). Nevertheless, the dependence of the theory upon it  
does not. We provide an argument to show that the choice of the K\"ahler potential does not affect 
the physics, but we are convinced that the implementation of local supersymmetry invariance coupling it to supergravity, 
might clarify this issue. 

The plan of the paper is the following: in Section \ref{bosonlag} we review the derivation of NS eqs. for the purely bosonic model. In particular, in \ref{bosact} two different methods to compute them are compared: the divergenceless condition on the energy-momentum tensor, and the invariance of the action under certain isometries; in \ref{boscl} Clebsch parametrization of the vector field $a_\m$ is considered.
In section \ref{complete} the supersymmetric completion of the previous model is taken into account, the action is constructed and explicit results are given for the bosonic sector. 
In \ref{susyclebsch} the supersymmetric generalization of the Clebsch parametrization is built, and the coupling to the linear multiplet $J$ is written.
In \ref{kpot} the issue of Ka\"hler potential and its appearance in the equations of motion is discussed.
Finally, in Appendix \ref{lagfinale} the complete supersymmetric Lagrangian is presented.

\section{Bosonic Lagrangian}
\label{bosonlag}

\subsection{Action and Equations of Motion}\label{bosact}
We first discuss the bosonic Lagrangian and we derive the equations of motion. The model is characterized 
by a divergenceless current $j^\mu$ and it is coupled to a worldvolume metric $g_{\mu\nu}$. In addition, 
we introduce an auxiliary gauge field $a_\mu$. The gauge invariance under $a_\mu \rightarrow a_\mu + \p_\mu \lambda$ 
is guaranteed by the conservation of $j^\mu$. The model is considered in 4d. There are two ways to get the equations of motion: 
the first one is by computing the energy-momentum tensor $T^{\mu\nu}$ and  requiring the vanishing of its divergence. The second method is 
requiring the invariance under certain isometries as will be discussed later. 

Let the action be
\begin{equation}
  \label{bola}
  {\mathcal{L}} = \sqrt{-g} \left( j^\mu a_\mu + f(j^2)\right)\,, \quad \quad   j^2 = j^\mu j^\nu g_{\mu\nu}\,.
\end{equation}
Note that the equation of motion obtained by taking the functional derivative w.r.t. an unconstrained 
$a_\mu$ yields $j^\mu=0$. Therefore, the correct equations of motion are obtained as follow: 
varying w.r.t. $j^\mu$ and $g_{\mu\nu}$ leads to 
\begin{equation}
a_\mu = -2 f'(j^2) j_\mu\,, 
\quad \quad 
T^{\mu\nu} = f'(j^2) \left( j^\mu j^\nu - g^{\mu\nu} j^2 \right) + \frac{1}{2} f(j^2) g^{\mu\nu}\,, 
\label{deltag}
\end{equation}
and the vanishing of the divergence of energy-momentum tensor implies
\begin{equation}
  \label{cons}
  \partial^\mu T_{\mu\nu} = 0 \longrightarrow  j^\mu [ f''(j^2) \left( j_\mu \partial_\nu j^2 - j_\nu \partial_\mu j^2 \right) + f'(j^2) \left( \partial_\nu j_\mu - \partial_\mu j_\nu \right) ] = 0\,.
\end{equation}
These are the usual NS equations which, together with the current $j^\mu$, yield the complete information on the fluid dynamics. 

Since we are primarily interested into $AdS/CFT$ correspondence, we recall that the fluid on the dual side must be a conformal one. That forces $f\left( j^{2} \right)$ to be equal to $C \left( j^{2} \right)^{2/3}$, where $C$ is a constant. This can be obtained by imposing the tracelessness of $T^{\mu\nu}$ or by studying the dilatation properties of the action, assuming that $j^{\mu}$ has dimension $3$ in 4d.

Notice that equation (\ref{cons}) can also be obtained in the following way: consider the field-strength associated to the abelian vector $a_\mu$,
$F_{\mu\nu} =  \left( \partial_\mu a_\nu - \partial_\nu a_\mu\right)$;
using the first of (\ref{deltag}) into $F$ and upon contraction with $j^\mu$ we get
\begin{equation}\label{EQMO}
j^\mu F_{\mu\nu} = \partial^\mu T_{\mu\nu} = 0 \,.
\end{equation}
It should be notice that, in both ways, the auxiliary field $a_\mu$ drops off the equations. 

Equation (\ref{EQMO}) calls for an explanation. First of all, we observe that, being $j^\mu$ a divergenceless current, 
action (\ref{bola}) is invariant under the gauge symmetry $\delta a_\mu = \partial_\mu \lambda$. We perform an isometry transformation which leaves the current $j^\mu$ invariant. 
In the form language, given ${\cal A} = a_\mu dx^\mu$, ${\cal J} = j^\mu \p_\mu$ and ${\cal X} = X^\mu \p_\mu$, we have
\begin{equation}\label{isozero}
{\cal L}_{\cal X}( {\cal A} ) = \iota_{\cal X} d {\cal A} + d ( \iota_{\cal X} {\cal A} )\,, 
\quad
{\cal L}_{\cal X}( {\cal J} ) = \Big[{\cal X}, {\cal J}\Big]=0\,,
\quad
 \end{equation}
$$
{\cal L}_{\cal X} ( g) = (\nabla_\mu X_\nu + \nabla_\nu X_\mu) dx^m \otimes dx^\n\,,
$$
 and in components 
\begin{equation}\label{iso}
\delta a_\mu = -F_{\mu\nu} X^\nu + \partial_{\mu} (a_\nu X^\nu)\,, \quad 
\delta j_\mu =0\,, \quad
\end{equation} 
$$
\delta g_{\mu\nu} = g_{\mu\rho}\partial_\nu X^\rho + g_{\nu\rho}\partial_\nu X^\rho + X^\rho \p_\rho g_{\m\n}=0 \,,
$$
where $X^\mu$ are the components of the Killing vector generating the isometry commuting 
with the current ${\cal J}$. 
Requiring the invariance of the action under such an isometry, one gets eqs.   (\ref{EQMO}). 

The condition 
$\delta j^\mu =0$ (if  $g_{\m\n} = \eta_{\m\n}$) 
can be reformulated as follows: given the vector field $X = X^\m \partial_\m$, the infinitesimal variation of $j^\mu$ can be expressed as
\begin{equation}\label{AWb}
\delta j^\mu = X^\n \partial_\n j^\m - j^\n \partial_\n X^\m  \,, \quad
\end{equation}
where the first term is a traslation parametrized by the coefficients $X^\nu$ and second term 
is a rotation with the parameter $\Lambda_{\m\n} = \half (\p_\m X_\n - \p_\n X_\m)$ due to 
Killing equation in (\ref{iso}). Condition (\ref{AWb}) can be rewritten as follows
\begin{equation}\label{AWc}
\Delta_X  j^\mu\equiv X^\n \p_\n j^\m = \Lambda^\m_{~\rho} \, j^\rho\,,
\end{equation}
which implies that the translation of the current $j^\m$ is compensated by a rotation. 
In the same way, the variation of $a_\mu$ can be cast in the form 
\begin{equation}\label{AWd}
\delta a_\mu = \Delta_X a_\mu + R_\mu^{~\nu} a_\nu \equiv 
 X^\n \p_\n a_\mu + \Lambda_{\mu}^{~\rho} a_\rho\,. 
\end{equation}
 Then, computing the variation of the action under a translation, we  have 
 \begin{eqnarray}\label{AWe}
\Delta_X S &=& \int  \Big( \Delta_X j^\mu a_\mu + j^\mu \Delta_X a_\mu + \Delta_X f(j^2)\Big)\nonumber \\
&=& \int  \Big( \Lambda_\mu^{~\nu} j^\mu a_\nu + j^\mu X^\nu \p_\nu a_\mu \Big)
\nonumber \\
&=& \int \Big( j^\mu \delta a_\mu \Big)
\nonumber \\
&=& \int  \Big(  j^\mu \left(- F_{\mu\nu} X^\nu + \p_\mu (a_\rho X^\rho) \right)\Big)\,. 
\end{eqnarray}
In the first line we have used eq.~(\ref{AWc}) and the Lorentz invariance of $f(j^2)$. From the second line 
to the third line, we have used the definition of the variation of the gauge potential $a_\mu$ under isometry (\ref{iso}) combined with a gauge variation. 
Thus, the second term vanishes because $j^\mu$ is divergenceless and from the first term, 
comparing with the definition of the energy-momentum tensor obtained by the N\"other theorem 
$\Delta_X S = \int  X^\mu \partial^\nu \, T_{\mu\nu}$, it yields 
\begin{equation}\label{AWf}
j^\mu F_{\mu\nu} = \partial^\mu T_{\mu\nu} =0\,.
\end{equation}
As a consistency condition, we must have $j^\nu \partial^\mu T_{\mu\nu} =0$, which can be easily verified using its explicit form (\ref{cons}). 

\subsection{Clebsch Parametrization of $a_\mu$}\label{boscl}

One may wonder why we adopt the above derivation of NS equations instead of computing directly the equations of motion by functional derivatives. 
Actually, it is possible to obtain them by means of variational principles, considering the auxiliary field $a_\m$ as parametrized by a set of potentials.
Moreover, since we would like to avoid any non-trivial solution for $a_\m$, we impose the constraint 
\begin{equation}\label{CSa}
F\wedge F =0\,, 
\end{equation}
where $F = d\, {\cal A}$ which, in components, becomes 
$\e^{\mu\n\rho\sigma} F_{\mu\n} F_{\rho\sigma} =0$. This constraint is 
equivalent to ${\cal A} \wedge F = d \Omega$ where $\Omega$ is a generic 2-form. 
It can be easily shown \cite{Zyoshi:2010} that the most general solution in 4d to (\ref{CSa}) is 
\begin{equation}\label{CSb}
{\cal A} = d\lambda + \alpha\, d \beta\,,
\end{equation}
where $\lambda, \a$ and $\b$ are zero forms. This implies that $F = d\alpha \wedge d\beta$ and 
the constraints (\ref{CSa}) follows immediately. This means that out of the four components of $a_\m$ only 3 of them survive the constraint and inserting them in the Lagrangian ({\ref{bola}}) 
we get 
\begin{equation}
  \label{bolalaa}
  {\mathcal{L}} =\Big( j^\mu (\p_\mu\lambda + \alpha \p_\mu \beta) + f(j^2)\Big) \,.
\end{equation}
The equations of motion are 
\begin{eqnarray}\label{bolalab}
\partial_\mu j^\m&=&0\,, \nonumber \\
j^\mu \partial_\mu \b &=&0\,, \nonumber \\
j^\mu \partial_\mu \a &=&0\,, \nonumber \\
\p_\mu\lambda + \alpha \p_\mu \beta + 2 j_\mu f'(j^2)&=&0\,.
\end{eqnarray}
With simple algebraic manipulations, one derives NS equations (\ref{EQMO}). 

 There is another way to parameterize the solution of (\ref{CSa}). Introducing one complex field $\phi$ and a 
 real function $K(\phi, \bar\phi)$, consequently $a_\mu$ becomes
 \begin{equation}\label{CSd}
a_\mu = \p_\mu \lambda + i (\partial K \partial_\mu \phi - \bar\partial K \partial_\mu \bar\phi)\,.
\end{equation}
If $K$ is identified with a K\"ahler potential for the complex manifold spanned by $\phi$, the second term in 
$a_\mu$ is the K\"ahler connection. Computing the field strength $F$ we get 
\begin{equation}\label{CSe}
F = - 2 i\, \partial \bar\partial K d \phi \wedge d\bar\phi\,.
\end{equation}
Namely, the manifold is a Hodge manifold where the $U(1)$ connection is related to the canonical 
$2$-form of the complex manifold. By the Bianchi identity, it follows that the canonical $2$-form 
$2 i\, \partial \bar\partial K d \phi \wedge d\bar\phi$, must be closed and therefore the space is K\"ahler. Notice that for a 
one dimensional complex manifold, no constraint on $K$ is due to its closure. 

The two parametrizations (\ref{CSb}) and (\ref{CSd}) are equivalent. This can be verified by 
assuming that $\alpha$ and $\beta$ are real functions of $\phi$ and $\bar\phi$. It yields 
\begin{equation}\label{CSf}
\alpha \partial \beta = i \partial K\,, \quad\quad
\alpha \bar\partial \beta = - i \bar\partial K\,.
\end{equation}
By dividing both equations by $\alpha$ and by computing the derivative we get 
\begin{equation}\label{CSg}
2 \partial \bar\partial K = (\partial K \bar\partial +  \bar\partial K \partial) \ln \alpha\,.
\end{equation}
This equation can be brought to quadrature. 
For example, assuming that $\alpha$ and $K$  are functions of the modulus $|\phi|^2 $, one can easily bring the above equation to an integral form. If 
$K(\phi, \bar\phi) = |\phi|^2$, then we get $\alpha = |\phi|^2$ and $\beta = i \ln (\phi/ \bar \phi)$. On the other hand, if $K(\phi, \bar\phi) = \ln (1 +|\phi|^2)$, then we get 
$\alpha =|\phi|^2/ (1+ |\phi|^2)$ and $\beta = i \ln (\phi/ \bar \phi)$. See also \cite{Jackiw:2004nm} for a 
discussion on this point.


\section{Supersymmetric Lagrangian}
\label{complete}

\subsection{Superfields, Action and Superfield Expansion}

We are now ready for the supersymmetrized version of the Lagrangian. We first construct the action in order to  
reproduce usual bosonic action (\ref{bola}) in the limit in which the fermions and the additional bosonic field are set to zero. 
A conserved current is a component of a linear multiplet in 4d and therefore we introduce a superfield $J$ for it. 
The auxiliary field $a_\mu$ is a component of the vector multiplet and we introduce a real superfield $A$.
Again we face with the problem of deriving the equations of motion since the superfield $A$ is constrained and, 
for that, we adopt a Clebsch parameterization. In the present case, it becomes natural to identify the abelian 
real superfield $A$ with a K\"ahler potential \cite{Nyawelo:2003bv} which is a real function of a chiral superfield $\phi$. 

$J$ and $A$ are defined as follows\footnote{In the following we use Weinberg notation
\cite{Weinberg:2000cr}. Nevertheless, we recall that in the language of \cite{superspace} a linear superfield is defined as $D^2 J = 0$ and $\bar D^2 \bar J =0$. If $J$ is a real linear superfield, $\bar J = J$ then the second condition follows from the first one.} 
\begin{equation}\label{LS}
\bar D D J = 0\,,  \hspace{2cm} \bar A = A\,,
\end{equation}
where $D = - \frac{\partial}{\partial \bar \theta} - (\gamma^\mu \theta) \partial_\mu$ and 
$\bar D = \frac{\partial}{\partial \theta} + (\gamma^\mu \bar\theta) \partial_\mu$  are the superderivatives. 
Using a linear superfield $J$, we automatically implement the conservation of the current $j^\mu$ which  is 
its $\theta^2$ component. The component expansion is given by 
\begin{equation}
J = C - i \bar \theta \gamma_5 \omega + \frac{i}{2} \bar \theta \gamma_5 \gamma_\mu \theta j^\mu + \frac{i}{2} \bar \theta \gamma_5 \theta \bar \theta \gamma^\mu \partial_\mu \omega + \frac{1}{8} (\bar \theta \gamma_5 \theta)^2 \square C,\label{eq:Jdef}
\end{equation}
and for the real superfield  in the Wess-Zumino gauge 
\begin{equation}
A = \frac{i}{2} \bar \theta \gamma_5 \gamma^\mu \theta a_\mu - i \bar \theta \gamma_5 \theta \bar \theta \lambda - \frac{1}{4} (\bar \theta \gamma_5 \theta)^2 D.\label{eq:Adef}
\end{equation}
The linear superfield contains one constrained vector $j^\m$, one scalar field $C$ and one Majorana spinor $\omega$. 
The vector can be dualized as $j^\mu = \epsilon^{\mu\nu\rho\sigma} H_{\nu\rho\sigma}$ where 
$H_{\mu\nu\rho}$ is the field strength of a 2-form potential $B_{\mu\nu}$. The latter can be further dualized into a scalar and therefore the linear multiplet has the same d.o.f. of an on-shell Wess-Zumino multiplet. 

Supersymmetry transformations are given by $\delta \Phi = \bar\alpha Q \Phi$ or, in component
\begin{align}\label{susy}
\delta j^\mu &= - \bar \alpha \gamma^{\mu\nu} \partial_\nu \omega\,, & \delta a_\mu &= \bar \alpha \gamma_\mu \lambda\,,  \nonumber\\
\delta \omega &= \left( - i \gamma_5 \gamma^\mu \partial_\mu C + \gamma_\mu j^\mu \right) \alpha\,, &
\delta \lambda &= - \left( i \, D \gamma_5 + F_{\mu\nu} \gamma^{\mu\nu} \right) \alpha\,,  \nonumber \\
\delta C &= i \, \bar \alpha \gamma_5 \omega\,, & \delta D &= i \, \bar \alpha \gamma_5 \gamma^\mu \partial_\mu \lambda\,.  \nonumber
\end{align}

Using the properties listed in the Appendix A it is possible to show that
\begin{equation}
\int d^4 x \int d^4 \theta [-JA] = \int d^4 x [j^\mu a_\mu + \bar \omega \lambda - CD]\,,
\end{equation}
which is the supersymmetric generalization of (\ref{bola}). In order to reproduce 
also the second term in (\ref{bola}), we need to introduce a new superfield defined as
\begin{equation}\label{newsuss}
{\cal J}_\mu =  \frac{1}{4 \, i}  (\bar D \gamma_5 \gamma_\mu D)J \,,
\end{equation}
which contains $j^\mu$ as the first component and its expansion is 
\begin{eqnarray}\label{newEXA} 
{\cal J}_\mu = \frac{1}{4 \, i}  (\bar D \gamma_5 \gamma_\mu D)J &=& j_\mu + \bar\theta \gamma_{\mu\nu} \partial^\n \om - \frac{i}{2} \bar\theta \g_5 \g^\n \theta \left( \p_\m \p_\n C - g_{\m\n} \square C \right) +\nonumber \\
&-& \half \bar\theta \g_5 \theta \bar\theta \g_5 \g^\nu (g_{\mu\n} \square \om - \p_\m \p_\n \omega) + \frac{1}{8} (\bar\theta \g_5 \theta)^2 \square j_\mu\,.
\end{eqnarray}
It should be noted that all terms in the above expansion are divergenceless. This can also be 
proven directly by the $D$-algebra and because of the linearity of the superfield $J$. Moreover, 
the new superfield ${\cal J}_\mu$ is itself a linear superfield. This can be seen by observing that 
each component of the superfield ${\cal J}_\mu$ is in the same relation with higher terms of the expansion 
as the components of the superfield $J$, and it can be checked by direct use of superderivatives.
 
Therefore the complete supersymmetric action is given by 
\begin{equation}\label{susyLAG}
S = \int d^4x \int d^4 \theta \Big( - J A +  F({\cal J}_\mu {\cal J}^\mu) \, J^2  \Big) \,.
\end{equation}
The minus sign in front of the first term is choosen to reproduce the normalization of the bosonic Lagrangian. 
The coefficients are chosen in order that (\ref{susyLAG}) coincides with the normalization of the bosonic Lagrangian where $f(j^2) = F(j^2) j^2$. 
 The argument of $F$, namely ${\cal J}_\mu {\cal J}^\mu$, is a dimensionful superfield and therefore it would be convenient to rescale it by a 
 dimensionful parameter. In the following, we will discard that parameter and we set it to 1. 

As discussed above, we would like to deal with superconformal fluid. 
For that, we require that the theory is conformal and supersymmetric, thus superconformal invariance follows. 
In particular, we first impose the dilatation properties of $F$ and it turns out  that $F\left( x \right)=C x^{-1/3}$.
That guarantees the conformal invariance of the action. The superconformal transformation rules for $J$ are deduced by its geometrical properties.

 To compute the component action, we need the expansion of 
 ${\cal J}_\mu {\cal J}^\mu$ and, using (\ref{newEXA}) we get 
 \begin{eqnarray}\label{EXAcalJ2}
{\cal J}_\mu {\cal J}^\mu 
	&=& j^2 
	+ 2 \, \bar\theta j_\m \g^{\m\n} \partial_\n \omega 
        + \bar \theta \theta \left( - \half \p_\m \bar \om \g^{\m\n} \p_\n \om - \frac{3}{4} \p_\m \bar \om \p^\m \om \right) + \nonumber \\
	&+& (\bar\theta \g_5 \theta) \left( - \half \p_\m \bar \om \g_5 \g^{\m\n} \p_\n \om - \frac{3}{4} \p_\m \bar \om \g_5 \p^\m \om \right) + \nonumber \\
        &+& \bar\theta \g_5 \g^\m \theta \left( i j_\m \square C - i j\cdot \p \p_\m C + \p_\m \bar \om \g_5 \not\!\p \om - \frac{1}{4} \p^\n \bar \om \g_5 \g_\m \p_\n \om \right) + \nonumber \\
	&+& (\bar\theta \g_5 \theta) \bar\theta \g_5 \left( j \cdot \p \not\!\p \om - j \cdot \g \square \om \right) + \bar\theta \g_5 \theta \bar\theta \left(2 i \not\!\p \omega \square C + i \g^\m \p^\n \om \p_\m \p_\n C \right) \nonumber \\   
&+& \frac{1}{4} (\bar\theta \g_5 \theta)^2 \left( j_\m \square j^\m + \p_\m \p_\n C \p^\m \p^\n C + 2 \square C \square C + \right. \nonumber \\
&+& \left. \p_\m \bar \omega \p^\m \omega + \p^\m \bar \omega \g_{\m\n} \p^\n \omega -2 \square \bar \omega \not\!\p \om \right) \,, 
\end{eqnarray}
similarly, for $J^2$, we have
 \begin{eqnarray}\label{EXAJ2}
J^2 &=& C^2 - 2 \, i \, C \bar\theta \g_5 \om + \nonumber \\
&+& \frac{1}{4}(\bar\theta \theta) \bar\om\om + \frac{1}{4} (\bar\theta  \g_5 \theta) \bar\om  \g_5 \om +  \bar\theta \g_5 \g_\m \theta ( i \, C j^\mu  + \frac{1}{4} \bar\om \g_5 \g^\m \om)  + \nonumber \\
&+& i \, \bar\theta  \g_5 \theta \bar\theta\!\!\not\!\partial \omega C + \,  \bar\theta  \g_5 \g_\mu \theta \bar\theta  \g_5 \om j^\m \nonumber \\
&+& \frac{1}{4} (\bar\theta \g_5 \theta)^2 ( C \square C + j^2 - \bar\om \!\!\not\! \p\om) \,. 
 \end{eqnarray}

Notice that the choice $f(j^2) = F(j^2) j^2$ does not spoil the generality of (\ref{susyLAG}) 
since it coincides with bosonic Lagrangian if $f(j^2)$ is defined up to an unessential constant. 
Action (\ref{susyLAG}) is chosen such that, by setting $C$ and $\omega$ to zero, it exactly reproduces the bosonic 
Lagrangian (\ref{bola}) and the corresponding NS equations. The presence of two different superfields, namely $J$ and 
${\cal J}_\mu$ in the Lagrangian is needed because of dimensional reasons or, equivalently, because $J$ does not start with $j^\mu$. 

In components the supersymmetric Lagrangian turns out to be
\begin{align}
  \int d^4 x \left[j^\mu a_\mu + \bar \omega \lambda - CD + \int d^4 \theta J^2  \sum_{i=0}^4 \frac{1}{i!}F^{(i)}(j^2) ({\cal J}_\mu 
  {\cal J}^\mu - j^2)^i  \right]\,, \label{ssl}
\end{align}
where we expanded the function $F$ around the first bosonic component of ${\cal J}_\mu {\cal J}^\mu$. 
The first term in the expansion reproduces the bosonic Lagrangian, while the other terms 
are classified according to their dimensions. Notice that the computation of the component action 
is made unhandy by the fact that there is a product of two or more superfields $({\cal J}_\mu 
  {\cal J}^\mu - j^2)^i J^2$. After the $\theta$-expansion is taken, one needs to compute all Fierz identities to simplify 
  the expressions and, finally, the integration over the $\theta$ variables can be taken. 

The first two terms in the expansion are
\begin{align}
&  \int d^4\theta \int d^4 x  \left[ F^{(0)}(j^2) J^2  + F^{(1)}(j^2) ({\cal J}_\mu {\cal J}^\mu - j^2) J^2  \right] = \nonumber\\
& = \int d^4 x  \Big\{\left[ F^{(0)}(j^2) (C \square C + j_\mu j^\mu - \bar\omega \gamma_\mu \p^\mu \omega) \right] + \nonumber \\
& + \Big[ F^{(1)}(j^2) 
\Big( 
 - C^2 [ j_\m \square j^\m + (\p_\m \p_\n C \p^\m\p^\n C + 2 \square C \square C )]  + 4 C j^\m j^\n \left(\p_\m \p_\n - \eta_{\m\n} \square \right) C + \nonumber\\
& - C^2  \p_\m \bar \omega \p^\m {\not\!\p} \omega 
+ 2 C^2 \square \bar \omega {\not\!\p} \omega - 2 i C j^\m \p_\m \bar \omega \g_5 {\not\!\p} \omega  - i C j^\m \p_\n \bar \omega \g_5 \g_\m \p^\n \omega + 4 C \square C \bar \omega {\not\!\p} \omega + \nonumber\\
& +2 C \p_\m\p_\n C \bar \omega \g^\m \p^\n \omega + 2 C j_\m \p_\n \bar \omega \g_\rho \p_\s \omega \varepsilon^{\m\n\rho\s}  - 2 i C j^\m \bar \omega \g_5 \p_\m {\not\!\p} \omega  + 2 i C j^\m \bar \omega \g_5 \g_\m \square \omega  
+ 2 j^2 \bar \omega {\not\!\p} \omega 
+ \nonumber\\
&- 2 j^\m j^\n \bar \omega \g_\m \p_\n \omega - i j^\m \left(\p_\m \p_\n - \eta_{\m\n} \square \right) C \bar \omega \g_5 \g^\n \omega 
- \frac{3}{4} \bar \omega \omega \p_\m \bar \omega \p^\m \omega  - \frac{1}{2} \bar \omega \omega \p_\m \bar \omega \g^{\m\n} \p_\n \omega + \nonumber\\
&+ \frac{3}{4} \bar \omega \g_5 \omega \p_\m \bar \omega \g_5 \p^\m \omega 
+ \frac{1}{2} \bar \omega \g_5 \omega \p_\m \bar \omega \g_5 \g^{\m\n} \p_\n \omega - \bar \omega \g_5 \g^\m \omega \p_\m \bar \omega \g_5 {\not\!\p} \omega 
+ \frac{1}{4} \bar \omega \g_5 \g_\m \omega \p_\n \bar \omega \g_5 \g^\m \p^\n \omega
 \Big) \Big]\Big\} \,.\label{18June1}
 \end{align}
As can be seen from this expression, these terms contain the interaction between the current $j^\mu$ and the 
fields $C$ and $\omega$. The part proportional to $F^{(1)}$ contains terms with four fields $\omega$ and therefore their 
self-interactions. In the forthcoming section, we will discuss the implications of those terms. Even though the action might seem 
bulky, it is a good starting point for the perturbation theory since the expansion is done in terms of higher derivative terms. 

Since the resulting action is rather cumbersome, we find convenient also to provide its bosonic truncation 
\begin{align}\label{bostrunch}
& \int d^4 x \left[ j^\mu a_\mu - CD + F^{(0)}(j^2) (C \square C + j_\mu j^\mu) \right] + \nonumber \\
& + \int d^4 x  \Big[ F^{(1)}(j^2) \Big( - C^2 j_\m \square j^\m - C^2 \left( \p_\m \p_\n - g_{\m\n} \square \right) C \left(\p^\m \p^\n - g^{\m\n} \square \right) C + \nonumber \\
& + 4 C j^\m j^\n \left(\p_\m \p_\n - g_{\m\n} \square \right) C \Big) \Big] + \nonumber \\
& + \int d^4 x  \Big[ F^{(2)}(j^2) \Big( - 4 C^2 j^\m j^\n \left( \p_\m \p_\rho - g_{\m\rho} \square \right) C \left(\p_\n \p^\rho - \d_\n^\rho \square \right) C \Big) \Big]\,.
  \end{align}
The bosonic action truncates at the second order in $F$, since all other terms are purely fermionic. This is due to the fact that 
in the expansion of the third power and of the fourth power, only those terms with a single $\theta$ contribute to the expansion since 
we have decided to expand around $j^\mu$. This simplifies the derivation of the energy-momentum tensor for the bosonic sector as we are going to 
discuss in the forthcoming section. In appendix B, all other terms are given. 


\subsection{Clebsch Parametrization for the Supersymmetric Case}
\label{susyclebsch}

We discuss here the Clebsch parameterization for the supersymmetric case. Here, 
the gauge field $a_\mu$ is replaced by the real superfield 
$A$ and therefore we have to parametrize it using a Clebsch parametrization as above. 
As suggested in \cite{Binetruy:2000zx} and in \cite{Nyawelo:2001fk}  we identify  
\begin{equation}\label{suCLEa}
A = \chi + \bar \chi + K(\phi, \bar \phi)\,,
\end{equation}
where $\chi, \phi$ and $\bar\chi, \bar\phi$ are chiral and anti-chiral fields, respectively. $K(\phi, \bar\phi)$ 
is a K\"ahler potential represented by a real function of the superfields $\phi$ and $\bar\phi$. 
The condition to be K\"ahler is $d {\cal K}=0$ where ${\cal K}$ is the canonical 2-form 
of the complex manifold spanned by $\phi$ and $\bar\phi$. Since the complex manifold is one dimensional, 
no interesting condition emerges from this constraint. 

The identification in (\ref{suCLEa}) implies that the Fayet-Ilioupoulos term induced by the abelian gauge field $A$ 
is given by 
\begin{equation}
S_{F-I} = \int d^4x d^4\theta A = \int d^4x d^4\theta K(\phi, \bar\phi) \,,
\end{equation}
which generates the dynamical equations of motion for the chiral fields (see for example \cite{Binetruy:2000zx}). 
In our case, however, this term is replaced by 
\begin{equation}\label{suCLEb}
S = \int d^4x d^4\theta (-J A + \dots)  = \int d^4x d^4\theta (-J (K(\phi,\bar \phi) + \chi + \bar \chi) + \dots)\,. 
\end{equation}
So, a fundamental difference is the absence of a naive kinetic term for $\phi$ and $\bar\phi$, but they are replaced by the 
superfield expansion of $JK$. The chiral field $\bar\chi$ and $\chi$ implement the linearity condition on $J$. 

Let us now consider the first term of action (\ref{suCLEb}) which, after Berezin integration reads 

\begin{align}
S = \int d^4 x &\frac{1}{2} K(\varphi , \bar \varphi) \square C + \nonumber \\
& - \p K \left(i j^\m \p_\m \varphi + \frac{1}{2} C \square \varphi
- i \frac{\sqrt{2}}{2} \bar \psi_L {\not\!\p} \lambda 
+ i \frac{\sqrt{2}}{2} \bar \lambda {\not\!\p} \psi_L \right) + c.c. +
\nonumber\\
& - \frac{1}{2} \p^2 K \left( C \p_\m \varphi \p^\m \varphi
- \sqrt{2} i \p_\m \varphi \bar \psi_L \g^\m \lambda  \right) + c.c. +
\nonumber\\
&- \p \bar \p K \left( 2 |P|^2 C
- C \p_\m \varphi \p^\m \bar \varphi 
+ \sqrt{2} i P \bar \psi_R \lambda  
- \sqrt{2} i \bar P \bar \psi_{L} \lambda 
+ \right. \nonumber \\
&\left. - C \bar \psi_{L} {\not\!\p} \psi_{R}
- C \bar \psi_{R} {\not\!\p} \psi_{L}
+ i j^\m \bar \psi_{L} \g_\m \psi_{R} 
+ \frac{\sqrt{2}}{2} i \p_\m \bar \varphi \bar \psi_{L} \g^\m \lambda
- \frac{\sqrt{2}}{2} i \p_\m \varphi  \bar\psi_{R} \g^\m \lambda  \right)
\nonumber\\
&- \frac{1}{3} \p^2 \bar \p K \left( 
- 2 C \bar P \bar \psi_L \psi_{L}
+ 2 C \p_\m \varphi \bar \psi_{L} \g^\m \psi_{R} 
- \sqrt{2} i \bar \psi_{L} \psi_{L} \bar \psi_{R} \lambda  \right) + c.c. +
\nonumber\\
&- \frac{1}{2} \p^2 \bar \p^2 K C \bar \psi_{R} \psi_{R} \bar \psi_{L} \psi_{L}\,,
\end{align}
where the chiral and antichiral superfields $\phi$ (respectively $\bar\phi$) are defined by condition
\begin{align}
&\frac{1-\g_5}{2} D \phi = 0\,, &\frac{1+\g_5}{2} D \bar\phi = 0\,,
\end{align}
and its components include a left-chiral spinor field $\psi_L = \left( \frac{1 + \g_5}{2} \right) \psi$ (respectively right-chiral $\psi_R$) and two scalar complex fields $\varphi$ and $P$ (respectively $\bar \varphi$ and $\bar P$).
The expression $Q_\m \equiv i \left( \p K \p_\m \varphi - \bar \p K \p_\m \bar \varphi \right) - i \p \bar \p K \bar \psi_{L} \g_\m \psi_{R} $ is known as the K\"ahler connection. Action (\ref{susyLAG}) contains a piece which depends upon the superfield $A$. Inserting the above expressions into (\ref{ssl}), we get 
an action which depends upon the components $\varphi, \psi_L$ and $F$ of the superfield $\phi$ (and its conjugated). Differentiation w.r.t. those fields, leads to the equations of motion. Truncating the action to its bosonic part, the first term in (\ref{susyLAG}) reads
\begin{align}
S = \int d^4 x & \left[ \frac{1}{2} K \square C - i j^\m \left( \p K \p_\m \varphi - \bar \p K \p_\m \bar \varphi \right) - \frac{1}{2} C \left( \p K \square \varphi + \bar \p K \square \bar \varphi \right) + \right. \nonumber \\
& - \left. \frac{1}{2} C \left( \p^2 K \p_\m \varphi \p^\m \varphi + \bar \p^2 K \p_\m \bar \varphi \p^\m \bar \varphi \right) - C \p \bar \p K \left( 2 |P|^2 - \p_\m \varphi \p^\m \bar \varphi \right) \right]\,,
\end{align}
where the K\"ahler potential $K$ is evaluated on $\varphi$ and its conjugated. 
Notice that, if we integrate by parts $K \square C$, the above expression simplifies considerably and becomes
\begin{align}
S = \int d^4 x \left[ i j^\m \left( \bar \p K \p_\m \varphi - \p K \p_\m \bar \varphi \right) - 2 C \p \bar \p K \left( |P|^2 - \p_\m \varphi \p^\m \bar \varphi \right) \right]\,.
\end{align}
The Lagrangian is diagonal in the auxiliary fields $P$ and their equations of motion (at the lowest level) imply either $C=0$ (fluid dynamics approximation) or 
$P= \bar P=0$ (which is the supersymmetric dynamics). 

To compute the equation of motion for the action, we recall that the expansion of $F$ is given in \ref{bostrunch}.
Varying w.r.t. $\varphi$ we get
\begin{equation}
	\begin{split}
		2\partial\bar\partial K \partial_{\mu} \bar\varphi\left( 
		i j^{\mu}-\partial^{\mu}C
		\right)
		-
		2C\partial\bar\partial K 
		\square \bar\varphi
		-
		2C\partial\bar\partial K
		\partial_{\mu}\bar\varphi\partial^{\mu}\bar\varphi
		=
		0
		\,.
	\end{split}
	\label{eomvarphi1}
\end{equation}
Analogously, we can get the equation of motion for $\bar\varphi$. For $j^{\mu}$ it reads
\begin{equation}
	\begin{split}
	&
	i\left( 
	\bar\partial K \partial_{\mu}\bar\varphi
	-\partial K \partial_{\mu}\varphi
	\right)
	+
	2F^{\left( 1 \right)} j^{\mu}
	\left( 
	C\square C +j^{2}
	\right)
	+2F j_{\mu}+
	\\&
	-
	2F^{\left( 2 \right)} C^{2} j_{\mu} j_{\nu}\square j^{\nu}
	-
	\square\left( 
	F^{\left( 1 \right)} C^{2} j_{\mu}
	\right)
	+\\&
	-
	F^{\left( 1 \right)}
	C^{2} \square j_{\mu}
	+8 F^{\left( 1 \right)} j^{\nu}\left( 
	\partial_{\mu}\partial_{\nu}
	-g_{\mu\nu}\square
	\right) C
	-8 F^{\left( 3 \right)}C^{2} j_{\mu}
	+\\&
	-4F^{\left( 2 \right)} C^{2} j^{\nu}\left( 
	\partial_{\mu}\partial_{\rho}
	-g_{\mu\rho}\square
	\right) C
	\left( \partial_{\nu}\partial^{\rho} 
	-\delta_{\nu}^{\rho}\square
	\right)C
	= 0
	\,,
	\end{split}
	\label{eomjmu}
\end{equation}
last, the one for $C$ is
\begin{equation}
	\begin{split}
		&
		2\partial\bar\partial K \partial_{\mu} \varphi \partial^{\mu}\bar\varphi
		+
		F\square C
		+
		\square F C
		-
		2F^{\left( 1 \right)} C j_{\mu}\square j^{\mu}
		+
		\\&
		+
		2 F^{\left( 1 \right)} \partial_{\mu}\partial_{\nu} C \partial^{\mu} \partial^{\nu} C
		+
		2
		\partial_{\mu}\partial_{\nu} \left( 
		F^{\left( 1 \right)}C^{2} \partial^{\mu} \partial^{\nu} C
		\right)
		+
		\\&
		+
		4F^{\left( 1 \right)} C \left( \square  C \right)^{2}
		+
		4\square \left( 
		F^{\left( 1 \right)}C^{2}\square C
		\right)
		+4 F^{\left( 1 \right)} j^{\mu} j^{\nu} 
		\left( \partial_{\mu}\partial_{\nu} -g_{\mu\nu} \square \right)
		 C
		+
		\\&
		+
		4 
		\left( \partial_{\mu}\partial_{\nu} -g_{\mu\nu} \square \right)
		\left( 
		F^{\left( 1 \right)} j^{\mu}j^{\nu} C
		\right) 
		+
		\\&
		-
		4 F^{\left( 2 \right)} C j^{\mu} j^{\nu} 
		\left( \partial_{\mu}\partial_{\rho} -g_{\mu\rho} \square \right)
		C 
		\left( \partial_{\nu}\partial^{\rho} -\delta_{\nu}^{\rho} \square \right)
		C
		+
		\\
		&
		-
		4 
		\left( \partial_{\mu}\partial_{\rho} -g_{\mu\rho} \square \right)
		\left( 
		F^{\left( 2 \right)} C^{2} j^{\mu} j^{\nu} 
		\left( \partial_{\nu}\partial^{\rho} -\delta_{\nu}^{\rho} \square \right)
		C
		\right)
		=
		0
		\,.
	\end{split}
	\label{eomC}
\end{equation}


\subsection{Superfield Equations}

Action (\ref{susyLAG}) is written in terms of 
a linear superfield $J$ and a real superfield $A$. For those superfields, the usual 
functional derivative cannot be used and therefore we cannot obtain the equations of motion 
by usual means (see \cite{superspace} for a complete discussion). 
To overcome such a problem, we add two auxiliary generic superfields $Z, S^\mu$, one chiral superfield $\chi$ 
and one antichiral superfield $\bar\chi$. 

The following action
\begin{eqnarray}\label{SFa}
S &=& \int d^4xd^4\theta \left( -J ( A + \chi + \bar \chi) \Big) +  
F\Big[ \Big(\frac{1}{4i}(\bar D \gamma_5 \gamma_\mu D) J\Big)^2\Big] J^2 \right)
\nonumber \\
&=& \int d^4xd^4\theta \left( -J ( A + \chi + \bar \chi) \Big) +
F[ {\cal J}^2] J^2 + 
S^\mu \Big[ \frac{1}{4i} (\bar D \gamma_5 \gamma_\mu D) J - {\cal J}_\mu\Big]  \right),
\end{eqnarray}
turns out to be equivalent to (\ref{susyLAG}). The chiral and antichiral superfields $\chi, \bar\chi$ impose the linearity condition on the superfield $J$. 

As already discussed above, in order to get the correct equations of motion, we replace the superfield $A$ with the K\"ahler 
potential. Then, we have 
\begin{equation}\label{KactA}
S_K= \int d^4xd^4\theta \left( - J ( K(\phi, \bar\phi) + \chi + \bar \chi) + 
F[ {\cal J}^2] J^2 + S^\mu \Big[ \frac{1}{4i} (\bar D \gamma_5 \gamma_\mu D) J - {\cal J}_\mu\Big]  \right)\,,
\end{equation}
from which we can get the equations of motion by taking the functional (constrained) derivatives with 
respect to superfields $J, \phi, \bar\phi, S^\mu, \chi, \bar\chi$ to get
\begin{eqnarray}\label{KactB}
\bar D D J &=&0\,, \nonumber \\
{\cal J}_\mu - \frac{1}{4i} (\bar D \gamma_5 \gamma_\mu D) J &=&0\,, \nonumber \\
S^\mu + 2 {\cal J}^\mu F'[ {\cal J}^2] J^2 &=&0\,, \nonumber \\
\bar D D( J \frac{\partial K}{\partial \phi} ) &=& 0\,, \nonumber \\
\bar D D( J \frac{\partial K}{\partial \bar\phi} ) &=& 0\,, \nonumber \\
K(\phi, \bar\phi) + \chi +\bar \chi - 2 J\, F[{\cal J}^2] - \frac{1}{4i} (\bar D \gamma_5 \gamma_\mu D) S^\mu&=&0\,.
\end{eqnarray}
 
 To study the above equations, we proceed as follows. The first eq. in (\ref{KactB}) implies the 
linearity of $J$ (and therefore its $\theta$ expansion is given by (\ref{eq:Jdef})). Then, we plug $J$ into the 
second equation for computing the vector superfield ${\cal J}_\mu$. Subsequently, we plug ${\cal J}_\mu$ 
into the third equation to evaluate $S^\mu$ and finally, by all those results, we can compute the value 
of $K$ in terms of the superfields $\phi$ and $\bar\phi$. 
Given that, eqs. (\ref{KactB}) become the new NS equations  only written in terms 
of the linear superfield $J$ which contains the physical degrees of freedom of our 
super-fluid.  

\subsection{Bosonic Sector}

In the present section, we study the model by setting to zero the fermions. We first derive the Lagrangian as a function 
of the fields $j^\mu$ and $C$ and then we provide a  new Lagrangian with new auxiliary fields which simplifies the 
derivation of the energy-momentum tensor.  

The bosonic part of the Lagrangian is (up to a factor $\sqrt{-g}$)
\begin{equation}\label{24JuneLbos}
\begin{split}
	\mathcal{L}_{bos}
	&=
	j^{\mu}a_{\mu} - C\,D + 
	\\&
	+ F^{(0)}\left( j^{2} \right)\left( C\square C +j_{\mu}j^{\mu}\right) +
	\\&+F^{(1)}\left( j^{2} \right)
	\left[ 
	-C^{2}j_{\mu}\square j^{\mu}
	+ C^{2}\partial_{\mu}\partial_{\nu}C\partial^{\mu}\partial^{\nu}C\right.
	\left.+2C^{2}\left( \square C \right)^{2}
	+4j^{\mu}j^{\nu} C \left( \partial_{\mu}\partial_{\nu}-g_{\mu\nu}\square \right)C
	\right]+
	\\&+
	\frac{1}{2}F^{(2)}\left( j^{2} \right)
	\left[ 
	-4C^{2}j^{\mu}j^{\nu}\left(  \partial_{\mu}\partial_{\rho}-g_{\mu\rho}\square \right)C
	\left(  \partial^{\rho}\partial_{\nu}-\delta^\rho_{\nu}\square \right)C
	\right]\,.
\end{split}
	\end{equation}
We define the quadratic differential operator
\begin{eqnarray}
	M_{\mu\nu} =  \partial_{\mu}\partial_{\nu}-g_{\mu\nu}\square\,, \quad\quad 
	\partial^\mu M_{\mu\nu} = 0\,, \quad\quad \square = - \frac{1}{3} g^{\mu\nu} M_{\mu\nu}\,. 
	\label{24JuneMmunu}
\end{eqnarray}
and we rewrite (\ref{24JuneLbos}) with the Lagrangian multiplier $S^{\mu\nu}$
\begin{equation}
	\begin{split}
		\mathcal{L}_{bos}
	&=
	j^{\mu}a_{\mu} - C\,D+
	\\&+
	F^{(0)}\left( j^{2} \right)\left( 
	-\frac{1}{3}g^{\mu\nu}B_{\mu\nu} C +j_{\mu}j^{\mu}
	\right)+ 
	\\&
	+F^{(1)}\left( j^{2} \right)
	\left[ 
	-C^{2}j_{\mu}\square j^{\mu}
	+ C^{2} B_{\mu\nu}B_{\rho\sigma}g^{\mu\rho}g^{\nu\sigma}
	+4 C j^{\mu}j^{\nu}B_{\mu\nu}
	\right]+
	\\&+
	\frac{1}{2}F^{(2)}\left( j^{2} \right)
	\left[ 
	-4C^{2}j^{\mu}j^{\nu}B_{\mu\rho}B_{\nu\sigma} g^{\rho\sigma}
	\right]+
	\\&+
	S^{\mu\nu}\left( 
	B_{\mu\nu}-M_{\mu\nu}C
	\right)
	\,.
	\end{split}
	\label{24JuneLbos2}
\end{equation}
In this way, we confine the covariantization of the differential operator $M_{\mu\nu}$ in a single term and the derivation 
of the energy-momentum tensor is greatly simplified. 
We now compute the equations of motion for $C$, $B_{\mu\nu}$ and $j^{\mu}$ respectively
\begin{equation}
	\begin{split}
	D=&
	-2F^{(1)}C+
	2F^{(1)}B_{\mu\nu}B^{\mu\nu}
	+
	4 F^{(1)}j^{\mu}j^{\nu}B_{\mu\nu}
	+\\&-
	4F^{(2)} Cj^{\mu}j^{\nu}B_{\mu\rho}B_{\nu \sigma}g^{\rho\sigma}
	-
	\frac{1}{3}F^{(0)}g^{\mu\nu}B_{\mu\nu}
	-
	M_{\mu\nu}S^{\mu\nu}\,,
	\end{split}
	\label{24JuneEomD}
\end{equation}
\begin{equation}
	\begin{split}
		S^{\mu\nu}
		=&
		-2F^{(1)} B^{\mu\nu} C^{2}
		-
		4C j^{\mu}j^{\nu}
		+
		4 F^{(2)} C^{2}j^{\mu}j^{\rho}B_{\rho\sigma}g^{\nu\sigma}
		+
		\frac{1}{3}F^{(0)} Cg^{\mu\nu}\,,
	\end{split}
	\label{24JuneEomB}
\end{equation}
\begin{equation}
	\begin{split}
		a_{\mu} =&
		-
		F^{(2)} j_{\mu}N_{\left[ 0 \right]}
		+
		F^{(1)} C^{2}\square j^{\mu}
		+\\&+
		\square\left( F^{(1)} C^{2}j^{\mu} \right)
		-
		8\,F^{(1)}\,C\,B_{\mu\nu}j^{\nu}
		-
		F^{(3)} N_{\left[ 1 \right]}
		+\\&+
		4F^{(2)}C^{2}B_{\mu\rho}B_{\nu\sigma}g^{\rho\sigma}j^{\nu}
		-
		2F^{(1)}j_{\mu}N_{\left[ 2 \right]}
		-	
		2F^{(0)}j_{\mu}\,,
	\end{split}
	\label{24JuneEomj}
\end{equation}
where $N_{\left[ 0 \right]}$, $N_{\left[ 1 \right]}$ and $N_{\left[ 2 \right]}$ are the terms in (\ref{24JuneLbos2}) proportional to $F^{(0)}$, $F^{(1)}$, and $F^{(2)}$, respectively.

In the case $j=0$, the Lagrangian (\ref{24JuneLbos2}) coupled to worldline metric is (we set $F^{(0)} = F^{(1)} = 1$)
\begin{equation}
	\mathcal{L}_{bos}{\big |}_{j=0} 
	=
	\sqrt{-g}\left[ 
	C^{2}
	g^{\mu\rho} g^{\nu\sigma}B_{\mu\nu}B^{\rho\sigma}
	-
	CD
	-
	\frac{1}{3} C g^{\mu\nu} B_{\mu\nu}
	+
	S^{\mu\nu}\left( B_{\mu\nu}-M_{\mu\nu} C \right)
	\right]\,.
	\label{24JLj0}
\end{equation}
The equations of motion for $C$ and $B_{\mu\nu}$ are
\begin{equation}
	D 
	=
	2C	g^{\mu\rho} g^{\nu\sigma}B_{\mu\nu}B^{\rho\sigma}
	-
	\frac{1}{3}g^{\mu\nu}B_{\mu\nu}
	-
	M_{\mu\nu}S^{\mu\nu}
	\,,
	\label{24JEoMCj0}
\end{equation}
and
\begin{equation}
	S^{\mu\nu}
	=
	-
	2C^{2}B^{\mu\nu}
	+
	\frac{1}{3}Cg^{\mu\nu}
	\,.
	\label{24JEoMBj0}
\end{equation}
Finally, for this simplified Lagrangian we derive the energy momentum tensor. We obtain
\begin{equation}
	\begin{split}
		T^{\mu\nu} = &
		-
		g^{\mu\nu} C\,\square C 
		-
		\frac{1}{2}g^{\mu\nu} \partial^{\rho}C\,\partial_{\rho} C
		+
		\partial^{\mu}C\,\partial^{\nu}C
		+\\&
		-
		\frac{5}{2} g^{\mu\nu} C^{2}\, \nabla^{\rho}\partial^{\sigma}C\,\nabla_{\rho}\partial_{\sigma}C
		-
		7g^{\mu\nu}C^{2}\,\square C\,\square C
		+\\&
		-
		2g^{\mu\nu}C\,\partial_{\rho}C\,\partial_{\sigma}C \nabla^{\rho}\partial^{\sigma}C
		-
		14
		g^{\mu\nu}C^{2}\,\partial_{\rho}C\,\partial^{\rho}\square C
		+\\&
		-
		3g^{\mu\nu} C^{3}\,\square\square C
		-
		8g^{\mu\nu}C\,\partial_{\rho}C\,\partial^{\rho}C\,\square C
		+\\&
		-
		C^{2}\,\square C\, \nabla^{\mu}\partial^{\nu}C
		+
		4C\,\partial_{\rho} C\, \nabla^{\rho}\partial^{\left( \mu \right.}C\,\partial^{\left.\nu \right)}C
		+\\&
		-2C\,\partial^{\rho} C\,\partial_{\rho}C\, \nabla^{\mu}\partial^{\nu}C
		+
		6C^{2}\,\partial^{\left( \mu \right.}C\,\nabla^{\left.\nu \right)}\square C
		+\\&
		-
		C^{2}\, \nabla^{\rho}\nabla^{\mu}\partial^{\nu}C\, \partial_{\rho}C
		+
		8C\,\partial^{\mu}C\,\partial^{\nu}C\,\square C
		\,.
	\end{split}
	\label{30JEnergyMomTensor}
\end{equation}

We prefer to analyze only the equations of motion with the Clebsch parametrization and the case $\omega=0$. This 
gives novel dynamical equations.  

\subsection{Dependence on the K\"ahler Potential}
\label{kpot}

We have to discuss the dependence of the equations of motion 
upon the K\"ahler potential. For that, we discuss only the bosonic sector 
and we observe the following identity 
\begin{equation}\label{depA}
-j^\mu F_{\mu\nu} + C \partial_\nu D = -4 \partial_\mu \left[ \partial \bar \partial K \, C ( \partial^\mu \bar\varphi  \partial_\nu \varphi  +  \partial^\mu \varphi  \partial_\nu \bar\varphi)\right]\,,
\end{equation}
where the r.h.s. can be also be written as $\partial_\mu (C G^{\mu\nu})$ where $G^{\mu\nu}$ is the inverse of the K\"ahler metric. It appears as a 
total derivative. However, we cannot discard such term. The reason is that it does not follows directly from the action, namely it is not a total derivative term 
derived from the action. Nevertheless, we can show that it is harmless and, at least in the rigid case, can be discarded. 

The left hand side of (\ref{depA}) can be obtained by the same method as in sec. (2.1). Indeed, by requiring the invariance under an isometry 
and using the same equations as above we get a new equation of the form 
\begin{equation}\label{depB}
\int d^4x X^\nu \left(-j^\mu F_{\mu\nu} + C \partial_\mu D \right)= 
- 4 \int d^4x X^\nu \partial_\mu \left[ \partial \bar\partial K \, C ( \partial^\mu \bar\varphi  \partial_\nu \varphi  +  \partial^\mu \varphi  \partial_\nu \bar\varphi)\right]\,.
\end{equation}
Now, we can use the integration by parts in the r.h.s. and by using the fact that $X^\mu$ must be a Killing vector for the 
flat metric we can easily conclude that the l.h.s. of (\ref{depA}) is effectively a total derivative and it can be discarded. A complete proof 
of this statement would be very interesting showing that the dynamical equations of motion are independent of the parametrization of 
the gauge field $A$. 

\section{Conclusions}\label{concl}

We propose a new supersymmetric action for supersymmetric fluid dynamics. We discuss several aspects such as the new NS equations and the derivation of them. A discussion on the Clebsch parametrization is proposed and the derivation of the superfield equations is done in that framework. There are several open issues: 1) what is the dynamics described by the present action? 2) what is the role of the boson $C$? 3) a fluid described only in terms of fermionic field 
can be discussed by setting to zero both $j^\mu$ and $C$. We believe that the 
study of the present system in the context of supergravity might shed some light on the coupling 
with the worldvolume metric and finally the susy partner of $T^{\mu\nu}$ can be computed. 
 We leave the discussion on supergravity to a forthcoming publication.

\section{Acknowledgments}
We are grateful to L. Andrianopoli, R. D'Auria and M. Trigiante for useful discussions. We also thank L. Gentile for the collaboration at the beginning of the present project and E. Ferrero for discussion on applications to ``real" physics. 


\appendix
\section{Fierz Identities}
\label{fierz}

We list here some of the properties of Majorana spinors and some useful Fierz Identities:

\begin{equation}
\label{symmetry}
\begin{split}
  \bar s_1 M s_2 &= \bar s_2 M s_1 \; \; {\rm{if}} \; M = 1,\gamma_5,
  \gamma_5 \gamma^\mu \,,\\
  \bar s_1 M s_2 &= - \bar s_2 M s_1 \; \; {\rm{if}} \; M =
  \gamma^\mu, \gamma^{\mu\nu}\,.
\end{split}
\end{equation}

The Fierz Identities for 2 identical spinors read

\begin{equation}
\label{fierzdue}
\theta \bar \theta = - \frac{1}{4} \left( \bar \theta \theta + \bar \theta \gamma_5 \theta \gamma_5 - \bar \theta \gamma_5 \gamma_\mu \theta \gamma_5 \gamma^\mu \right)\,,
\end{equation}
while those for 3 spinors are

\begin{equation}
\label{fierztre}
\begin{split}
  \theta ( \bar \theta \theta) &= - \gamma_5 \theta \bar \theta \gamma_5 \theta \,,\\
  \theta ( \bar \theta \gamma_5 \gamma_\mu \theta) &= - \gamma_\mu \theta \bar \theta \gamma_5 \theta \,.
\end{split}
\end{equation}

Using (\ref{fierztre}) it is easy to show that the following identities also hold

\begin{equation}
\label{tetaquattro}
\begin{split}
  (\bar \theta \theta)^2 &= - (\bar \theta \gamma_5 \theta)^2 \,,\\
  (\bar \theta \gamma_5 \gamma_\mu \theta) (\bar \theta \gamma_5 \gamma_\nu \theta) &= - \eta_{\mu\nu} (\bar \theta \gamma_5 \theta)^2 \,,\\
  (\bar \theta \theta) (\bar \theta \gamma_5 \theta) &= (\bar \theta \theta) (\bar \theta \gamma_5 \gamma_\mu \theta) = (\bar \theta \gamma_5 \theta) (\bar \theta \gamma_5 \gamma_\mu \theta) = 0 \,.
\end{split}
\end{equation}

Finally the integration measure for Grassmann variables is

\begin{equation}
\int d^4 \theta (\bar \theta \gamma_5 \theta)^2 = -4 \label{berezin}\,.
\end{equation}


\section{Complete Lagrangian}
\label{lagfinale}

Here we present the complete expansion of the supersymetric Lagrangian (\ref{susyLAG}). This can be rewritten as
\begin{equation}
	\begin{split}
		\mathcal{L}=
		\int d^4 x\int d^{4}\theta\left( 
		-
		J A
		+		
		\sum_{i=0}^{4}\frac{1}{i!}
		F^{\left( i \right)}L_{i}
		\right)\,,
\end{split}
	\label{AppsusyLagExp1}
\end{equation}
where $F^{\left( i \right)}$ is the order $i$-derivative of $ F({\cal J}_\mu {\cal J}^\mu)$ computed at $ {\cal J}_\mu {\cal J}^\mu = j_{\mu}j^{\mu}$ and
\begin{equation}
	\begin{split}
		L_{i}=&\left(  {\cal J}_\mu {\cal J}^\mu -j_{\mu}j^{\mu} \right)^{i}J^{2}\,.
	\end{split}
	\label{AppdefLi}
\end{equation}
In the following we show the explicit form of the four $L_{i}$. To perform the computation we developed a program written in FORM language (see \cite{Vermaseren:2000nd} and references therein) which, given a set of superfields expanded in components, returns as result any desired combination of these fields, integrated over $d^4 \theta$. The 
subroutine structure of the program allows us to check every intermediate passage, or to use each single procedure to perform different calculations such as Fierz identities or gamma manipulations. 

Notice that only $L_{1}$ and $L_{2}$ has purely bosonic terms (\ref{L1bos}) and (\ref{L2bos}).

\begin{subequations}
\begin{align}
L_{1} = &
- C^2 \left[ j_\m \square j^\m + \left(\p_\m \p_\n C \p^\m\p^\n C + 2 \square C \square C \right) \right] + \nonumber\\
& + 4 C j^\m j^\n \left(\p_\m \p_\n - \eta_{\m\n} \square \right) C  + \label{L1bos}\\
& - C^2  \p_\m \bar \omega \p^\m {\not\!\p} \omega + \nonumber\\
& + 2 C^2 \square \bar \omega {\not\!\p} \omega + \nonumber\\
& - 2 i C j^\m \p_\m \bar \omega \g_5 {\not\!\p} \omega + \nonumber\\
& - i C j^\m \p_\n \bar \omega \g_5 \g_\m \p^\n \omega + \nonumber\\
& + 4 C \square C \bar \omega {\not\!\p} \omega + \nonumber\\
& + 2 C \p_\m\p_\n C \bar \omega \g^\m \p^\n \omega + \nonumber\\
& + 2 C j_\m \p_\n \bar \omega \g_\rho \p_\s \omega \varepsilon^{\m\n\rho\s} + \nonumber\\
& - 2 i C j^\m \bar \omega \g_5 \p_\m {\not\!\p} \omega  + \nonumber\\
& + 2 i C j^\m \bar \omega \g_5 \g_\m \square \omega + \nonumber\\
& + 2 j^2 \bar \omega {\not\!\p} \omega + \nonumber\\
& - 2 j^\m j^\n \bar \omega \g_\m \p_\n \omega + \nonumber\\
& - i j^\m \left(\p_\m \p_\n - \eta_{\m\n} \square \right) C \bar \omega \g_5 \g^\n \omega + \nonumber\\
& - \frac{3}{4} \bar \omega \omega \p_\m \bar \omega \p^\m \omega + \nonumber\\
& - \frac{1}{2} \bar \omega \omega \p_\m \bar \omega \g^{\m\n} \p_\n \omega + \nonumber\\
& + \frac{3}{4} \bar \omega \g_5 \omega \p_\m \bar \omega \g_5 \p^\m \omega + \nonumber\\
& + \frac{1}{2} \bar \omega \g_5 \omega \p_\m \bar \omega \g_5 \g^{\m\n} \p_\n \omega + \nonumber\\
& - \bar \omega \g_5 \g^\m \omega \p_\m \bar \omega \g_5 {\not\!\p} \omega + \nonumber\\
& + \frac{1}{4} \bar \omega \g_5 \g_\m \omega \p_\n \bar \omega \g_5 \g^\m \p^\n \omega
\,,
\end{align}
\end{subequations}

\begin{subequations}
\begin{align}
L_{2} =& - 4 C^2 j^\m j^\n \left( \p_\m \p_\rho - \eta_{\m\rho} \square \right) C \left( \p_\n \p^\rho - \d_\n^\rho \square \right) C  
+ \label{L2bos}\\
& - 2 C^2 j^\m \left(\p_\m \p_\n - \eta_{\m\n} \square \right) C \left[ \p_\rho \bar \omega \g_\s \p_\tau \omega \varepsilon_{\n\rho\s\tau} - 6 i \p^\n \bar \omega \g_5 {\not\!\p} \omega + i \p_\rho \bar \omega \g_5 \g^\n \p^\rho \omega \right] \nonumber\\
& + 6 C^2 \square C j_\m \left[ \p_\n \bar \omega \g_\rho \p_\s \omega \varepsilon^{\m\n\rho\s} + 2 i  \p^\m \bar \omega \g_5 {\not\!\p} \omega - 2 i \p_\n \bar \omega \g_5 \g^\m \p^\n \omega \right] \nonumber\\
& + 2 C^2 j_\m \left( \p_\a \p_\n - \eta_{\a\n} \square \right) C \p_\rho \bar \omega \g^\a \p_\s \omega \varepsilon^{\m\n\rho\s} + \nonumber\\
& - 4 i C^2 \left( \p_\m \p_\n - \eta_{\m\n} \square \right) C j^\rho \p^\m \bar \omega \g_5 \g_\rho \p^\n \omega + \nonumber\\
& + \frac{9}{4} C^2 \p_\m \bar \omega \p^\m \omega \p_\n \bar \omega \p^\n \omega + \nonumber\\
& + 3 C^2 \p_\m \bar \omega \p^\m \omega \p_\n \bar \omega \g^{\n\rho} \p_\rho \omega + \nonumber\\
& - \frac{9}{4} C^2 \p_\m \bar \omega \g_5 \p^\m \omega \p_\n \bar \omega \g_5 \p^\n \omega + \nonumber\\
& - 3 C^2 \p_\m \bar \omega \g_5 \p^\m \omega \p_\n \bar \omega \g_5 \g^{\n\rho} \p_\rho \omega + \nonumber\\
& - 2 C^2 \p_\m \bar \omega \g_5 \g^\n \p^\m \omega \p_\n \bar \omega \g_5 {\not\!\p} \omega + \nonumber\\
& + \frac{1}{4} C^2 \p_\m \bar \omega \g_5 \g^\n \p^\m \omega \p_\rho \bar \omega \g_5 \g_\n \p^\rho \omega + \nonumber\\
& + 4 C^2 \p^\m \bar \omega \g_5 {\not\!\p} \omega \p_\m \bar \omega \g_5 {\not\!\p} \omega + \nonumber\\
& + C^2 \p_\m \bar \omega \g^{\m\n} \p_\n \omega \p_\rho \bar \omega \g^{\rho\s} \p_\s \omega + \nonumber\\
& - C^2 \p_\m \bar \omega \g_5 \g^{\m\n} \p_\n \omega \p_\rho \bar \omega \g_5 \g^{\rho\s} \p_\s \omega + \nonumber\\
& + 4 C^2 j^\m j^\n \p_\m \bar \omega \g_\n \square \omega + \nonumber\\
& - 4 C^2 j^\m j^\n \p_\m \p_\n \bar \omega {\not\!\p} \omega + \nonumber\\
& - 4 C^2 j^\m j^\n\p_\rho \bar \omega \g_\m \p_\n \p_\rho \omega + \nonumber\\
& + 4 C^2 j^2 \square \bar \omega {\not\!\p} \omega + \nonumber\\
& + 4 i C^2 j^\tau j_\m \p_\n \bar \omega \g_5 \g_\rho \p_\tau \p_\s \omega \varepsilon^{\m\n\rho\s} + \nonumber\\
& - 4 C j^2 \left[ j_\m \p_\n \bar \omega \g_\s \p_\rho \omega \varepsilon^{\m\n\rho\s} - 2 i j^\m \p_\m \bar \omega \g_5 {\not\!\p} \omega + 2i j^\n \p_\m \bar \omega \g_5 \g_\n \p^\m \omega \right] + \nonumber\\
& + 4 i C j^\m j^\n j^\rho \p_\m \bar \omega \g_5 \g_\n \p_\rho \omega + \nonumber\\
& - 8 i C j^2 j^\m \p_\m \bar \omega \g_5 {\not\!\p} \omega + \nonumber\\
& + 4 i C j^2 j^\m \p_\n \bar \omega \g_5 \g_\m \p^\n \omega + \nonumber\\
& - 8 C j^\m j^\n \left( \p_\m \p_\n - \eta_{\m\n} \square \right) C \bar\omega {\not\!\p} \omega + \nonumber\\
& + 8 C j^\m j^\n \left(\p_\n \p_\rho - \eta_{\n\rho} \square \right) C \bar\omega \g_\n \p^\rho \omega + \nonumber\\
& + 8 i C j^\tau j_\m \left(\p_\tau \p_\n - g_{\tau\n} \square \right) C \bar \omega \g_5 \g_\rho \p_\s \omega \varepsilon^{\m\n\rho\s} + \nonumber\\
& - 8 i C j^\n \bar \omega \g_\n \p_\m \omega \p^\m \bar \omega \g_5 {\not\!\p} \omega + \nonumber\\
& + 2 i C j^\rho \bar \omega \g_\rho \p_\m \omega \p_\n \bar \omega \g_5 \g^\m \p^\n\omega + \nonumber\\
& + 8 i C j^\m \bar\omega {\not\!\p} \omega \p_\m \bar \omega \g_5 {\not\!\p} \omega + \nonumber\\
& - 2 i C j^\m \bar \omega {\not\!\p} \omega \p_\n \bar \omega \g_5 \g_\m \p^\n \omega + \nonumber\\
& - 2 C j_\m \bar \omega \g_5 \g_\n \p_\rho \omega \p_\tau \bar \omega \g_5 \g_\s \p^\tau \omega \varepsilon^{\m\n\rho\s} + \nonumber\\
& + 8 C j_\m \bar \omega \g_5 \g_\n \p_\rho \omega \p_\s \bar\omega\g_5 {\not\!\p} \omega \varepsilon^{\m\n\rho\s}
+ \nonumber\\
& + 6 i C j_\m \bar \omega \g^{\m\n} \p_\n \omega \p_\rho \bar \omega \g_5 \p^\rho \omega + \nonumber\\
& + 4 i C j_\m \bar\omega\g^{\m\n} \p_\n \omega \p_\rho \bar \omega \g_5 \g^{\rho\s} \p_\s \omega + \nonumber\\
& - 6 i C j_\m \bar \omega \g_5 \g^{\m\n} \p_\n \omega \p_\rho \bar \omega \p^\rho \omega + \nonumber\\
& - 4 i C j_\m \bar \omega \g_5 \g^{\m\n} \p_\n \omega \p_\rho \bar \omega \g^{\rho\s} \p_\s \omega + \nonumber\\
& + \bar \omega \omega \left[ j^\m j^\n \p_\m \bar \omega \p_\n \omega - j^2 \p_\m \bar \omega \p^\m \omega + 2 j^\m j_\n \p_\m \bar \omega \g^{\n\rho} \p_\rho \omega - j^2 \p_\m \bar \omega \g^{\m\n} \p_\n \omega \right] + \nonumber\\
& + \bar \omega \g_5 \omega \left[ - j^\m j^\n \p_\m \bar \omega \g_5 \p_\n \omega + j^2 \p_\m \bar \omega \g_5 \p^\m \omega - 2 j^\m j_\n \p_\m \bar \omega \g_5 \g^{\n\rho} \p_\rho \omega + j^2 \p_\m \bar \omega \g_5 \g^{\m\n} \p_\n \omega \right] + \nonumber\\
& + j^\m \bar \omega \g_5 \g_\m \omega \left[ - i \varepsilon^{\n\rho\s\tau} j_\n \p_\rho \bar \omega \g_\s \p_\tau \omega + 2 j^\n \p_\n \bar \omega \g_5 {\not\!\p} \omega - 2 j^\n \p_\rho \bar \omega \g_5 \g_\n \p^\rho \omega \right] + \nonumber\\
& + \bar \omega \g_5 \g_\m \omega \left[ - i \varepsilon^{\m\n\rho\s} j^\tau j_\n \p_\rho \bar \omega \g_\tau \p_\s \omega - j^\n j^\rho \p_\n \bar \omega \g_5 \g^\m \p_\rho \omega + 2 j^\n j^\rho \p_\n \bar \omega \g_5 \g_\rho \p^\m \omega + \right. \nonumber\\
& \left. - 2 j^2 \p^\m \bar \omega \g_5 {\not\!\p} \omega + j^2\p_\n \bar \omega \g_5 \g^\m \p^\n \omega \right]
\,,
\end{align}
\end{subequations}

\begin{align}
	L_{3}
	=&
 +12 C^2 \left(\partial_{\mu}\partial_{\nu}-g_{\mu\nu}\square \right) C
 \left[ 
 j^{\mu}j^{\nu}j_{\tau}\partial_{\rho}\bar\omega\gamma_{\lambda}\partial_{\sigma}\omega\varepsilon^{\tau\rho\sigma\lambda}
 \,+ \right.
 \nonumber\nonumber\\&
 \left.
 -2i\ j^{\mu}j^{\nu}j^{\rho}\partial_{\rho}\bar\omega\gamma_{5}{\not\!\p}\omega
 +2i\ j^{\mu}j^{\nu}j^{\rho}\partial^{\sigma}\bar\omega\gamma_{5}\gamma_{\rho}\partial_{\sigma}\omega
 \,+ \right.
 \nonumber\nonumber\\&
 \left.
 +j^{\mu}j^{\rho}j^{\sigma}\partial_{\alpha}\bar\omega\gamma_{\rho}\partial_{\beta}\omega\varepsilon^{\sigma\nu\alpha\beta}
 +i\ j^{\mu}j^{\rho}j^{\sigma}\partial_{\rho}\bar\omega\gamma_{5}\gamma^{\nu}\partial_{\sigma}\omega
 \,+ \right.
 \nonumber\nonumber\\&
 \left.
 -2i\ j^{\mu}j^{\rho}j^{\sigma}\partial_{\rho}\bar\omega\gamma_{5}\gamma_{\sigma}\partial^{\nu}\omega
 +2i\ \left(j\cdot j\right) j^{\mu}\partial^{\nu}\bar\omega\gamma_{5}{\not\!\p}\omega
 \,+ \right.
 \nonumber\nonumber\\&
 \left.
 -i\ \left(j\cdot j\right)  j^{\mu}\partial^{\rho}\bar\omega\gamma_{5}\gamma^{\nu}\partial_{\rho}\omega
 \right]
+ \nonumber\\&-6 C^2j^{\nu}j^{\mu} \partial_{\mu}\bar\omega\partial_{\nu}\omega  \partial_{\rho}\bar\omega\gamma^{\rho\sigma}\partial_{\sigma}\omega \,   
 + \nonumber\\&-9 C^2j^{\mu}j^{\nu}\partial^{\rho} \bar\omega\partial_{\rho}\omega \partial_{\mu}\bar\omega\partial_{\nu}\omega  \,   
+ \nonumber\\&+9 C^2\left(j\cdot j\right) \partial^{\mu}\bar\omega\partial_{\mu}\omega \partial^{\nu}\bar\omega\partial_{\nu}\omega  \,   
+ \nonumber\\&-18 C^2j^{\nu}j_{\rho} \partial^{\mu}\bar\omega\partial_{\mu}\omega \partial_{\nu}\bar\omega\gamma^{\rho\sigma}\partial_{\sigma}\omega  \,   
+ \nonumber\\&-15 C^2\left(j\cdot j\right) \partial^{\mu}\bar\omega\partial_{\mu}\omega \partial_{\nu}\bar\omega\gamma^{\rho\nu}\partial_{\rho}\omega  \,   
+ \nonumber\\&-6 C^2j^{\mu}j^{\nu} \partial_{\mu}\bar\omega\gamma_{5}\partial_{\nu}\omega  \partial_{\rho}\bar\omega\gamma_{5}\gamma^{\sigma\rho}\partial_{\sigma}\omega \,   
+ \nonumber\\&+9 C^2j^{\mu}j^{\nu} \partial^{\rho}\bar\omega\gamma_{5}\partial_{\rho}\omega \partial_{\nu}\bar\omega\gamma_{5}\partial_{\rho}\omega  \,   
+ \nonumber\\&-9 C^2\left(j\cdot j\right) \partial^{\mu}\bar\omega\gamma_{5}\partial_{\mu}\omega \partial_{\nu}\bar\omega\gamma_{5}\partial_{\nu}\omega  \,   
+ \nonumber\\&+18 C^2j^{\nu}j_{\rho} \partial^{\mu}\bar\omega\gamma_{5}\partial_{\mu}\omega \partial_{\nu}\bar\omega\gamma_{5}\gamma^{\rho\sigma}\partial_{\sigma}\omega  \,   
+ \nonumber\\&-15 C^2\left(j\cdot j\right) \partial^{\mu}\bar\omega\gamma_{5}\partial_{\mu}\omega \partial_{\nu}\bar\omega\gamma_{5}\gamma^{\nu\rho}\partial_{\rho}\omega  \,   
+ \nonumber\\&+3 i C^2  j^{\mu}j_{\nu} \partial_{\alpha}\bar\omega\gamma_{\mu}\partial_{\rho}\omega \partial^{\sigma}\bar\omega\gamma_{5}\gamma_{\lambda}\partial_{\sigma}\omega \varepsilon^{\nu\alpha\rho\lambda}    \,   
+ \nonumber\\&-12 i C^2 j^{\mu}j_{\alpha} \partial_{\nu}\bar\omega\gamma_{\mu}\partial_{\rho}\omega \partial^{\sigma}\bar\omega\gamma_{5}\gamma_{\sigma}\partial_{\lambda}\omega \varepsilon^{\alpha\nu\rho\lambda}    \,   
+ \nonumber\\&-12 i C^2 j_{\mu}j^{\lambda} \partial_{\nu}\bar\omega\gamma_{\rho}\partial_{\sigma}\omega \partial_{\lambda}\bar\omega\gamma_{5}{\not\!\p}\omega  \varepsilon^{\mu\nu\sigma\rho}    \,   
+ \nonumber\\&+3 i C j^{\mu}j_{\nu} \partial_{\rho}\bar\omega\gamma_{\sigma}\partial_{\lambda}\omega \partial^{\alpha}\bar\omega\gamma_{5}\gamma_{\mu}\partial_{\alpha}\omega \varepsilon^{\nu\rho\lambda\sigma}   \,   
+ \nonumber\\&-6 C^2  j^{\mu} j^{\nu}  \partial_{\mu}\bar\omega\gamma_{5}\gamma_{\rho}\partial_{\rho}\omega\partial_{\nu}\bar\omega\gamma_{5}{\not\!\p}\omega \,   
+ \nonumber\\&-6 C^2 j^{\mu}j^{\nu} \partial_{\rho}\bar\omega\gamma_{5}\gamma_{\mu}\partial_{\nu}\omega \partial^{\sigma}\bar\omega\gamma_{5}\gamma_{\sigma}\partial_{\rho}\omega \,   
+ \nonumber\\&+3 C^2j^{\nu}j^{\rho}  \partial_{\mu}\bar\omega\gamma_{5}\gamma_{\nu}\partial_{\rho}\omega \partial^{\sigma}\bar\omega\gamma_{5}\gamma_{\mu}\partial_{\sigma}\omega \,   
+ \nonumber\\&-18 C^2 j^{\mu}j^{\nu} \partial_{\mu}\bar\omega\gamma_{5}{\not\!\p}\omega \partial_{\nu}\bar\omega\gamma_{5}{\not\!\p}\omega  \,   
+ \nonumber\\&+30 C^2j^{\mu} j^{\nu} \partial_{\mu}\bar\omega\gamma_{5}{\not\!\p}\omega  \partial^{\sigma}\bar\omega\gamma_{5}\gamma_{\nu}\partial_{\sigma}\omega \,   
+ \nonumber\\&-6 C^2j^{\mu}j^{\nu} \partial^{\rho}\bar\omega\gamma_{5}\gamma_{\mu}\partial_{\rho}\omega  \partial^{\sigma}\bar\omega\gamma_{5}\gamma_{\nu}\partial_{\sigma}\omega  \,   
+ \nonumber\\&-3 C^2j^{\mu}j^{\nu}  \partial^{\rho}\bar\omega\gamma_{5}\gamma^{\sigma}\partial_{\rho}\omega \partial_{\nu}\bar\omega\gamma_{5}\gamma_{\sigma}\partial_{\mu}\omega \,   
+ \nonumber\\&+3 C^2j^{\mu}j^{\nu} \partial^{\rho}\bar\omega\gamma_{5}\gamma^{\sigma}\partial_{\rho}\omega \partial_{\nu}\bar\omega\gamma_{5}\gamma_{\mu}\partial_{\sigma}\omega  \,   
+ \nonumber\\&-18 C^2\left(j\cdot j\right)  \partial^{\mu}\bar\omega\gamma_{5}\gamma^{\nu}\partial_{\mu}\omega \partial_{\nu}\bar\omega\gamma_{5}{\not\!\p}\omega \,   
+ \nonumber\\&+3 C^2 \left(j\cdot j\right) \partial^{\mu}\bar\omega\gamma_{5}\gamma^{\nu}\partial_{\mu}\omega \partial^{\rho}\bar\omega\gamma_{5}\gamma_{\nu}\partial_{\rho}\omega \,   
+ \nonumber\\&+12 C^2j^{\mu}j^{\nu}  \partial^{\sigma}\bar\omega\gamma_{5}{\not\!\p}\omega \partial_{\mu}\bar\omega\gamma_{5}\gamma_{\sigma}\partial_{\nu}\omega \,   
+ \nonumber\\&-6 C^2 j^{\mu}j^{\nu} \partial^{\sigma}\bar\omega\gamma_{5}{\not\!\p}\omega \partial_{\nu}\bar\omega\gamma_{5}\gamma_{\mu}\partial_{\sigma}\omega \,   
+ \nonumber\\&+24\left(j\cdot j\right)  C^2 \partial^{\mu}\bar\omega\gamma_{5}{\not\!\p}\omega \partial_{\mu}\bar\omega\gamma_{5}{\not\!\p}\omega \,   
+ \nonumber\\&-12 C^2 j^{\mu}j^{\nu} \partial^{\rho}\bar\omega\gamma_{5}{\not\!\p}\omega \partial_{\nu}\bar\omega\gamma_{5}\gamma_{\mu}\partial_{\rho}\omega  \,   
+ \nonumber\\&-12 C^2j_{\mu}j^{\nu} \partial_{\nu}\bar\omega\gamma^{\mu\rho}\partial_{\rho}\omega  \partial_{\sigma}\bar\omega\gamma^{\sigma\lambda}\partial_{\lambda}\omega \,   
+ \nonumber\\&+6 C^2  \left(j\cdot j\right) \partial_{\mu}\bar\omega\gamma^{\mu\nu}\partial_{\nu}\omega \partial_{\rho}\bar\omega\gamma^{\rho\sigma}\partial_{\sigma}\omega\,   
+ \nonumber\\&+12 C^2  j_{\mu}j^{\nu}\partial_{\nu}\bar\omega\gamma_{5}\gamma^{\mu\rho}\partial_{\rho}\omega \partial_{\sigma}\bar\omega\gamma_{5}\gamma^{\sigma\lambda}\partial_{\lambda}\omega \,   
+ \nonumber\\&-6 C^2\left(j\cdot j\right)  \partial_{\mu}\bar\omega\gamma_{5}\gamma^{\mu\nu}\partial_{\nu}\omega \partial_{\rho}\bar\omega\gamma_{5}\gamma^{\rho\sigma}\partial_{\sigma}\omega \,   
+ \nonumber\\&-4 C j^{\mu}j^{\nu} j_{\rho}\bar\omega\gamma_{\mu}\partial_{\nu}\omega  \partial_{\sigma}\bar\omega\gamma_{\beta}\partial_{\alpha}\omega \varepsilon^{\rho\sigma\alpha\beta} \,  \,   
+ \nonumber\\&+8 i C j^{\mu}j^{\nu} j^{\rho} \bar\omega\gamma_{\mu}\partial_{\nu}\omega  \partial_{\rho}\bar\omega\gamma_{5}{\not\!\p}\omega   \,  \,   
+ \nonumber\\&-8 i C j^{\mu}j^{\nu}j^{\rho}  \bar\omega\gamma_{\mu}\partial_{\nu}\omega  \partial^{\sigma}\bar\omega\gamma_{5}\gamma_{\rho}\partial_{\sigma}\omega    \,  \,   
+ \nonumber\\&-4 C j^{\mu}j^{\nu}j^{\alpha}  \bar\omega\gamma_{\mu}\partial_{\rho}\omega  \partial_{\sigma}\bar\omega\gamma_{\alpha}\partial_{\beta}\omega  \varepsilon^{\nu\rho\sigma\beta}\,  \,   
+ \nonumber\\&-4 i C j^{\mu}j^{\rho}j^{\nu} \bar\omega\gamma_{\mu}\partial^{\sigma}\omega  \partial_{\nu}\bar\omega\gamma_{5}\gamma_{\sigma}\partial_{\rho}\omega     \,  \,   
+ \nonumber\\&+8 i C j^{\mu}  j^{\rho}j^{\nu}  \bar\omega\gamma_{\mu}\partial^{\sigma}\omega \partial_{\nu}\bar\omega\gamma_{5}\gamma_{\rho}\partial_{\sigma}\omega   \,  \,   
+ \nonumber\\&-8 i C \left(j\cdot j\right) j^{\mu} \bar\omega\gamma_{\mu}\partial^{\nu}\omega  \partial_{\nu}\bar\omega\gamma_{5}{\not\!\p}\omega    \,  \,   
+ \nonumber\\&+4 i C \left(j\cdot j\right) j^{\mu}  \bar\omega\gamma_{\mu}\partial^{\nu}\omega \partial^{\rho}\bar\omega\gamma_{5}\gamma_{\nu}\partial_{\rho}\omega    \,  \,   
+ \nonumber\\&+4 C \left(j\cdot j\right) j_{\mu} \bar\omega{\not\!\p}\omega \partial_{\nu}\bar\omega\gamma_{\rho}\partial_{\lambda}\omega \varepsilon^{\mu\nu\lambda\rho} \,  \,   
+ \nonumber\\&-4 i C j^{\mu}j^{\nu}j^{\rho}  \bar\omega{\not\!\p}\omega \partial_{\rho}\bar\omega\gamma_{5}\gamma_{\mu}\partial_{\nu}\omega    \,  \,   
+ \nonumber\\&+4 i C \left(j\cdot j\right) j^{\mu} \bar\omega{\not\!\p}\omega \partial^{\rho}\bar\omega\gamma_{5}\gamma_{\mu}\partial_{\rho}\omega     \,  \,   
+ \nonumber\\&+8 i C  j^{\mu}  j^{\nu}j^{\rho}  \bar\omega\gamma_{5}\gamma^{\sigma}\partial_{\mu}\omega \partial_{\rho}\bar\omega\gamma_{\nu}\partial_{\sigma}\omega    \,  \,   
+ \nonumber\\&-8 i C  j^{\mu} j^{\nu}j^{\rho}  \bar\omega\gamma_{5}\gamma_{\mu}\partial^{\sigma}\omega \partial_{\rho}\bar\omega\gamma_{\nu}\partial_{\sigma}\omega    \,  \,   
+ \nonumber\\&-8 i C \left(j\cdot j\right) j^{\mu} \bar\omega\gamma_{5}\gamma_{\nu}\partial^{\rho}\omega \partial_{\rho}\bar\omega\gamma_{\mu}\partial_{\nu}\omega    \,  \,   
+ \nonumber\\&-4 C j^{\mu}j^{\nu} j_{\rho} \bar\omega\gamma_{5}\gamma_{\sigma}\partial_{\lambda}\omega \partial_{\nu}\bar\omega\gamma_{5}\gamma_{\alpha}\partial_{\mu}\omega  \varepsilon^{\rho\lambda\sigma\alpha}\,  \,   
+ \nonumber\\&+8 C j^{\mu}j^{\nu}  j_{\rho} \bar\omega\gamma_{5}\gamma_{\sigma}\partial_{\lambda}\omega \partial_{\nu}\bar\omega\gamma_{5}\gamma_{\mu}\partial_{\alpha}\omega \varepsilon^{\rho\lambda\sigma\alpha}\,  \,   
+ \nonumber\\&+4 C \left(j\cdot j\right) j_{\mu}\bar\omega\gamma_{5}\gamma_{\nu}\partial_{\rho}\omega \partial^{\sigma}\bar\omega\gamma_{5}\gamma_{\alpha}\partial_{\sigma}\omega \varepsilon^{\mu\rho\nu\alpha}  \,  \,   
+ \nonumber\\&-8 C\left(j\cdot j\right)j_{\mu} \bar\omega\gamma_{5}\gamma_{\nu}\partial_{\rho}\omega \partial_{\alpha}\bar\omega\gamma_{5}{\not\!\p}\omega \varepsilon^{\mu\rho\nu\alpha}  \,  \,   
+ \nonumber\\&-4 i C j_{\mu}j^{\nu}j^{\rho} \bar\omega\gamma^{\mu\sigma}\partial_{\sigma}\omega  \partial_{\rho}\bar\omega\gamma_{5}\partial_{\nu}\omega    \,  \,   
+ \nonumber\\&+4 i C \left(j\cdot j\right) j_{\mu}  \bar\omega\gamma^{\mu\nu}\partial_{\nu}\omega \partial^{\rho}\bar\omega\gamma_{5}\partial_{\rho}\omega     \,  \,   
+ \nonumber\\&-8 i C j_{\mu}  j_{\nu}j^{\rho} \bar\omega\gamma^{\mu\sigma}\partial_{\sigma}\omega \partial_{\rho}\bar\omega\gamma_{5}\gamma^{\nu\alpha}\partial_{\alpha}\omega     \,  \,  
+ \nonumber\\&+4 i C  \left(j\cdot j\right) j_{\mu}  \bar\omega\gamma^{\mu\nu}\partial_{\nu}\omega \partial_{\rho}\bar\omega\gamma_{5}\gamma^{\rho\sigma}\partial_{\sigma}\omega    \,  \,   
+ \nonumber\\&+4 i C j_{\mu}j^{\nu}j^{\rho}   \bar\omega\gamma_{5}\gamma^{\mu\sigma}\partial_{\sigma}\omega  \partial_{\nu}\bar\omega\partial_{\rho}\omega   \,  \,   
+ \nonumber\\&-4 i C \left(j\cdot j\right) j^{\mu}  \bar\omega\gamma_{5}\gamma_{\mu\nu}\partial_{\nu}\omega \partial^{\rho}\bar\omega\partial_{\rho}\omega    \,  \,   
+ \nonumber\\&+8 i C j_{\mu} j_{\nu}j^{\rho} \bar\omega\gamma_{5}\gamma^{\mu\sigma}\partial_{\sigma}\omega  \partial_{\rho}\bar\omega\gamma^{\nu\alpha}\partial_{\alpha}\omega     \,  \,   
+ \nonumber\\&-4 i C\left(j\cdot j\right) j_{\mu} \bar\omega\gamma_{5}\gamma^{\mu\nu}\partial_{\nu}\omega \partial_{\rho}\bar\omega\gamma^{\rho\sigma}\partial_{\sigma}\omega    \,,
\label{23Junefour}
\end{align}

\begin{align}
L_{4}
=
 &+4 C^2 j^{\mu}j^{\nu}j^{\rho}j^{\sigma} \partial_{\mu}\bar\omega\partial_{\nu}\omega \, \partial_{\rho}\bar\omega\partial_{\sigma}\omega  \,   
+ \nonumber\\&+16 C^2 j^{\mu}j^{\nu} j_{\rho}j^{\sigma} \partial_{\mu}\bar\omega\partial_{\nu}\omega  \partial_{\sigma}\bar\omega\gamma^{\rho\lambda}\partial_{\lambda}\omega \,   
+ \nonumber\\&-8 C^2  \left(j\cdot j\right)j^{\mu}j^{\nu}\partial_{\mu}\bar\omega\partial_{\nu}\omega  \partial_{\rho}\bar\omega\gamma^{\rho\sigma}\partial_{\sigma}\omega \,   
+ \nonumber\\&-8 C^2\left(j\cdot j\right) j^{\mu}j^{\nu}  \partial^{\rho}\bar\omega\partial_{\rho}\omega \partial_{\mu}\bar\omega\partial_{\nu}\omega  \,   
+ \nonumber\\&+4 C^2  \left(j\cdot j\right)^2  \partial^{\mu}\bar\omega\partial_{\mu}\omega \partial^{\nu}\bar\omega\partial_{\nu}\omega\,   
+ \nonumber\\&-16 C^2  \left(j\cdot j\right) j_{\mu}j^{\nu}\partial^{\rho}\bar\omega\partial_{\rho}\omega \partial_{\nu}\bar\omega\gamma^{\mu\sigma}\partial_{\sigma}\omega \,   
+ \nonumber\\&+8 C^2  \left(j\cdot j\right)^2\partial^{\mu}\bar\omega\partial_{\mu}\omega \partial_{\rho}\bar\omega\gamma^{\rho\sigma}\partial_{\sigma}\omega \,   
+ \nonumber\\&-4 C^2 j^{\mu}j^{\nu} j^{\rho}j^{\sigma} \partial_{\mu}\bar\omega\gamma_{5}\partial_{\nu}\omega  \partial_{\rho}\bar\omega\gamma_{5}\partial_{\sigma}\omega  \,   
+ \nonumber\\&-16 C^2 j^{\mu}j^{\nu}j_{\rho}j^{\sigma} \partial_{\mu}\bar\omega\gamma_{5}\partial_{\nu}\omega  \partial_{\sigma}\bar\omega\gamma_{5}\gamma^{\rho\lambda}\partial_{\lambda}\omega  \,   
+ \nonumber\\&+8 C^2 \left(j\cdot j\right)  j^{\mu}j^{\nu}\partial_{\mu}\bar\omega\gamma_{5}\partial_{\nu}\omega \partial_{\rho}\bar\omega\gamma_{5}\gamma^{\rho\sigma}\partial_{\sigma}\omega \, 
+ \nonumber\\&+8 C^2 \left(j\cdot j\right) j^{\mu}j^{\nu}\partial^{\rho}\bar\omega\gamma_{5}\partial_{\rho}\omega \partial_{\mu}\bar\omega\gamma_{5}\partial_{\nu}\omega  \,   
+ \nonumber\\&-4 C^2 \left(j\cdot j\right)^2 \partial^{\mu}\bar\omega\gamma_{5}\partial_{\mu}\omega \partial^{\nu}\bar\omega\gamma_{5}\partial_{\nu}\omega \,   
+ \nonumber\\&+16 C^2 \left(j\cdot j\right) j_{\mu}j^{\nu} \partial^{\rho}\bar\omega\gamma_{5}\partial_{\rho}\omega \partial_{\nu}\bar\omega\gamma_{5}\gamma^{\mu\sigma}\partial_{\sigma}\omega \,  
+ \nonumber\\&-8 C^2 \left(j\cdot j\right)^2 \partial^{\mu}\bar\omega\gamma_{5}\partial_{\mu}\omega \partial_{\nu}\bar\omega\gamma_{5}\gamma^{\nu\rho}\partial_{\rho}\omega  \,   
+ \nonumber\\&-4 C^2 \left(j\cdot j\right)j^{\mu}j^{\nu}\partial_{\mu}\bar\omega\gamma^{\rho}\partial^{\sigma}\omega  \partial_{\nu}\bar\omega\gamma_{\rho}\partial_{\sigma}\omega   \, 
+ \nonumber\\&-4 C^2 j^{\mu}j^{\nu}j^{\rho}j^{\sigma}\partial_{\mu}\bar\omega\gamma_{\nu}\partial^{\lambda}\omega  \partial_{\rho}\bar\omega\gamma_{\sigma}\partial_{\lambda}\omega  \,   
+ \nonumber\\&+4 C^2 \left(j\cdot j\right) j^{\mu} j^{\nu} \partial_{\mu}\bar\omega\gamma^{\rho}\partial^{\sigma}\omega \partial_{\nu}\bar\omega\gamma_{\sigma}\partial_{\rho}\omega \,   
+ \nonumber\\&+12 C^2  j^{\mu}j^{\nu}j^{\rho}j^{\sigma} \partial^{\lambda}\bar\omega\gamma_{\mu}\partial_{\nu}\omega \partial_{\rho}\bar\omega\gamma_{\sigma}\partial_{\lambda}\omega  \,   
+ \nonumber\\&+12 C^2 \left(j\cdot j\right) j^{\mu}  j^{\nu}  \partial^{\rho}\bar\omega\gamma^{\sigma}\partial_{\mu}\omega\partial_{\nu}\bar\omega\gamma_{\sigma}\partial_{\rho}\omega \,   
+ \nonumber\\&-12 C^2 \left(j\cdot j\right) j^{\mu}j^{\nu}   \partial^{\rho}\bar\omega\gamma^{\sigma}\partial_{\mu}\omega  \partial_{\nu}\bar\omega\gamma_{\rho}\partial_{\sigma}\omega \,   
+ \nonumber\\&-24 C^2 \left(j\cdot j\right) j^{\mu}j^{\nu}\partial^{\sigma}\bar\omega\gamma^{\rho}\partial_{\mu}\omega  \partial_{\rho}\bar\omega\gamma_{\nu}\partial_{\sigma}\omega  \,   
+ \nonumber\\&+8 C^2 \left(j\cdot j\right)j^{\mu}j^{\nu}  \partial^{\rho}\bar\omega\gamma_{\mu}\partial_{\sigma}\omega \partial_{\rho}\bar\omega\gamma^{\sigma}\partial_{\nu}\omega  \,   
+ \nonumber\\&-8 i C^2 j^{\mu}  j^{\nu}j^{\rho} j_{\sigma} \partial_{\alpha}\bar\omega\gamma_{\mu}\partial_{\beta}\omega \partial_{\rho}\bar\omega\gamma_{5}\gamma_{\lambda}\partial_{\nu}\omega\varepsilon^{\sigma\alpha\beta\lambda}    \,   
+ \nonumber\\&+16 i C^2 j^{\mu} j^{\nu}j^{\rho} j_{\sigma}  \partial_{\lambda}\bar\omega\gamma_{\mu}\partial_{\alpha}\omega  \partial_{\rho}\bar\omega\gamma_{5}\gamma_{\nu}\partial_{\beta}\omega\varepsilon^{\sigma\lambda\alpha\beta}   \,   
+ \nonumber\\&+8  iC^2 \left(j\cdot j\right)j^{\mu} j^{\nu}\partial_{\rho}\bar\omega\gamma_{\mu}\partial_{\sigma}\omega  \partial^{\lambda}\bar\omega\gamma_{5}\gamma_{\alpha}\partial_{\lambda}\omega \varepsilon^{\nu\rho\sigma\alpha}     \,   
+ \nonumber\\&-16 i C^2  \left(j\cdot j\right)  j^{\mu} j^{\nu}  \partial_{\rho}\bar\omega\gamma_{\mu}\partial_{\sigma}\omega\partial_{\alpha}\bar\omega\gamma_{5}{\not\!\p}\omega \varepsilon^{\nu\rho\sigma\alpha}    \,   
+ \nonumber\\&+8 C^2  \left(j\cdot j\right)^2  \partial^{\mu}\bar\omega\gamma^{\nu}\partial^{\rho}\omega \partial_{\mu}\bar\omega\gamma_{\nu}\partial_{\rho}\omega\,   
+ \nonumber\\&-16 C^2 \left(j\cdot j\right)^2  \partial^{\mu}\bar\omega\gamma^{\nu}\partial^{\rho}\omega \partial_{\mu}\bar\omega\gamma_{\rho}\partial_{\nu}\omega \,   
+ \nonumber\\&+8i C^2   j^{\mu}j^{\nu}j^{\rho}j^{\sigma} \partial_{\lambda}\bar\omega\gamma_{\alpha}\partial_{\beta}\omega \partial_{\rho}\bar\omega\gamma_{5}\gamma_{\mu}\partial_{\nu}\omega \varepsilon^{\sigma\lambda\beta\alpha}   \,   
+ \nonumber\\&-8i C^2  \left(j\cdot j\right)  j^{\mu} j^{\nu}\partial_{\rho}\bar\omega\gamma_{\sigma}\partial_{\lambda}\omega \partial^{\alpha}\bar\omega\gamma_{5}\gamma_{\mu}\partial_{\alpha}\omega \varepsilon^{\nu\rho\lambda\sigma}   \,   
+ \nonumber\\&+4 C^2  j^{\mu}j^{\nu} j^{\rho}j^{\sigma}\partial_{\nu}\bar\omega\gamma_{5}\gamma^{\lambda}\partial_{\mu}\omega \partial_{\sigma}\bar\omega\gamma_{5}\gamma_{\lambda}\partial_{\rho}\omega \,   
+ \nonumber\\&-8 C^2  j^{\mu}j^{\nu} j^{\rho}j^{\sigma} \partial_{\nu}\bar\omega\gamma_{5}\gamma_{\mu}\partial^{\lambda}\omega \partial_{\sigma}\bar\omega\gamma_{5}\gamma_{\lambda}\partial_{\rho}\omega\,   
+ \nonumber\\&+4 C^2 j^{\mu}j^{\nu} j^{\rho}j^{\sigma}\partial_{\nu}\bar\omega\gamma_{5}\gamma_{\mu}\partial^{\lambda}\omega  \partial_{\sigma}\bar\omega\gamma_{5}\gamma_{\rho}\partial_{\lambda}\omega \,   
+ \nonumber\\&+8 C^2j^{\mu}  j^{\nu}j^{\rho}j^{\sigma}\partial_{\mu}\bar\omega\gamma_{5}{\not\!\p}\omega  \partial_{\sigma}\bar\omega\gamma_{5}\gamma_{\nu}\partial_{\rho}\omega \,   
+ \nonumber\\&-4 C^2 \left(j\cdot j\right) j^{\mu}j^{\nu} \partial_{\mu}\bar\omega\gamma_{5}{\not\!\p}\omega  \partial_{\nu}\bar\omega\gamma_{5}{\not\!\p}\omega   \,   
+ \nonumber\\&-8 C^2 j^{\mu}j^{\nu} j^{\rho}j^{\sigma} \partial^{\lambda}\bar\omega\gamma_{5}\gamma_{\mu}\partial_{\nu}\omega \partial_{\sigma}\bar\omega\gamma_{5}\gamma_{\lambda}\partial_{\rho}\omega \,   
+ \nonumber\\&+12 C^2  j^{\mu}j^{\nu}j^{\rho}j^{\sigma}\partial^{\lambda}\bar\omega\gamma_{5}\gamma_{\mu}\partial_{\nu}\omega \partial_{\sigma}\bar\omega\gamma_{5}\gamma_{\rho}\partial_{\lambda}\omega  \,   
+ \nonumber\\&-8 C^2 \left(j\cdot j\right) j^{\mu}j^{\nu} \partial^{\rho}\bar\omega\gamma_{5}\gamma_{\mu}\partial_{\nu}\omega \partial_{\rho}\bar\omega\gamma_{5}{\not\!\p}\omega \,   
+ \nonumber\\&+8 C^2\left(j\cdot j\right)  j^{\mu}j^{\nu} \partial^{\rho}\bar\omega\gamma_{5}\gamma_{\mu}\partial_{\nu}\omega\partial^{\sigma}\bar\omega\gamma_{5}\gamma_{\rho}\partial_{\sigma}\omega  \,   
+ \nonumber\\&+8 C^2 j^{\mu} j^{\nu}j^{\rho}j^{\sigma}\partial_{\mu}\bar\omega\gamma_{5}{\not\!\p}\omega \partial_{\sigma}\bar\omega\gamma_{5}\gamma_{\nu}\partial_{\rho}\omega  \,   
+ \nonumber\\&-12 C^2\left(j\cdot j\right)j^{\mu}j^{\nu} \partial_{\mu}\bar\omega\gamma_{5}{\not\!\p}\omega  \partial_{\nu}\bar\omega\gamma_{5}{\not\!\p}\omega   \,   
+ \nonumber\\&+8 C^2 \left(j\cdot j\right)j^{\mu} j^{\nu}  \partial_{\mu}\bar\omega\gamma_{5}{\not\!\p}\omega \partial_{\sigma}\bar\omega\gamma_{5}\gamma_{\nu}\partial_{\sigma}\omega \,   
+ \nonumber\\&-16 C^2j^{\mu}j^{\nu}j^{\rho}j^{\sigma} \partial^{\lambda}\bar\omega\gamma_{5}\gamma_{\mu}\partial_{\lambda}\omega  \partial_{\sigma}\bar\omega\gamma_{5}\gamma_{\nu}\partial_{\rho}\omega  \,   
+ \nonumber\\&+8 C^2\left(j\cdot j\right)j^{\mu}j^{\nu}  \partial^{\rho}\bar\omega\gamma_{5}\gamma_{\mu}\partial_{\rho}\omega  \partial_{\nu}\bar\omega\gamma_{5}{\not\!\p}\omega  \,   
+ \nonumber\\&-8 C^2\left(j\cdot j\right) j^{\mu}j^{\nu}  \partial_{\mu}\bar\omega\gamma_{5}\gamma^{\rho}\partial_{\nu}\omega \partial^{\sigma}\bar\omega\gamma_{5}\gamma^{\rho}\partial_{\sigma}\omega \,   
+ \nonumber\\&+8 C^2\left(j\cdot j\right)j^{\mu}j^{\nu}   \partial^{\rho}\bar\omega\gamma_{5}\gamma^{\sigma}\partial_{\rho}\omega \partial_{\nu}\bar\omega\gamma_{5}\gamma_{\mu}\partial_{\sigma}\omega \,   
+ \nonumber\\&-16 C^2 \left(j\cdot j\right)^2\partial^{\mu}\bar\omega\gamma_{5}\gamma^{\nu}\partial_{\mu}\omega \partial_{\nu}\bar\omega\gamma_{5}{\not\!\p}\omega  \,   
+ \nonumber\\&+4 C^2 \left(j\cdot j\right)^2 \partial^{\mu}\bar\omega\gamma_{5}\gamma^{\nu}\partial_{\mu}\omega \partial^{\rho}\bar\omega\gamma_{5}\gamma_{\nu}\partial_{\rho}\omega  \,   
+ \nonumber\\&+16 C^2 \left(j\cdot j\right)j^{\mu}j^{\nu} \partial^{\sigma}\bar\omega\gamma_{5}{\not\!\p}\omega \partial_{\nu}\bar\omega\gamma_{5}\gamma_{\sigma}\partial_{\mu}\omega  \,   
+ \nonumber\\&-24 C^2\left(j\cdot j\right)  j^{\mu}j^{\nu} \partial^{\sigma}\bar\omega\gamma_{5}{\not\!\p}\omega \partial_{\nu}\bar\omega\gamma_{5}\gamma_{\mu}\partial_{\sigma}\omega \,   
+ \nonumber\\&+16 C^2 \left(j\cdot j\right)^2\partial^{\nu}\bar\omega\gamma_{5}{\not\!\p}\omega \partial_{\nu}\bar\omega\gamma_{5}{\not\!\p}\omega  \,   
+ \nonumber\\&+16 C^2j_{\mu}j^{\nu} j_{\rho}j^{\sigma} \partial_{\nu}\bar\omega\gamma^{\mu\lambda}\partial_{\lambda}\omega \partial_{\sigma}\bar\omega\gamma^{\rho\alpha}\partial_{\alpha}\omega  \,   
+ \nonumber\\&-16 C^2 \left(j\cdot j\right) j_{\mu}j^{\nu} \partial_{\nu}\bar\omega\gamma^{\mu\rho}\partial_{\rho}\omega  \partial_{\sigma}\bar\omega\gamma^{\sigma\lambda}\partial_{\lambda}\omega \,   
+ \nonumber\\&+4 C^2 \left(j\cdot j\right)^2 \partial_{\mu}\bar\omega\gamma^{\mu\nu}\partial_{\nu}\omega \partial_{\rho}\bar\omega\gamma^{\rho\sigma}\partial_{\sigma}\omega  \,   
+ \nonumber\\&-16 C^2 j_{\mu}j^{\nu}j_{\rho}j^{\sigma} \partial_{\nu}\bar\omega\gamma_{5}\gamma^{\mu\lambda}\partial_{\lambda}\omega  \partial_{\sigma}\bar\omega\gamma_{5}\gamma^{\rho\alpha}\partial_{\alpha}\omega  \,   
+ \nonumber\\&+16 C^2\left(j\cdot j\right)j^{\mu}j^{\nu} \partial_{\nu}\bar\omega\gamma_{5}\gamma^{\mu\rho}\partial_{\rho}\omega  \partial_{\sigma}\bar\omega\gamma_{5}\gamma^{\sigma\lambda}\partial_{\lambda}\omega  \,   
+ \nonumber\\&-4 C^2\left(j\cdot j\right)^2 \partial_{\mu}\bar\omega\gamma_{5}\gamma^{\mu\nu}\partial_{\nu}\omega \partial_{\rho}\bar\omega\gamma_{5}\gamma^{\rho\sigma}\partial_{\sigma}\omega  \,.
\label{23Junefive}
\end{align}


\end{document}